\documentclass[pre, twocolumn, preprintnumbers, amsmath, amssymb, nofootinbib, floatfix, superscriptaddress]{revtex4}

\usepackage{graphicx, bm, xcolor, bbold}
\usepackage{enumerate}
\usepackage{graphicx,bm}
\usepackage[normalem]{ulem} 
\usepackage[breaklinks]{hyperref}

\makeatletter

\def\graphicscale{\twocolumn@sw{0.4}{0.4}}
\def\graphicthreescale{\twocolumn@sw{0.3}{0.4}}

\begin{document}

\title{Out-of-equilibrium dynamics across the first-order quantum
  transitions\\ of one-dimensional quantum Ising models}

\author{Andrea Pelissetto}
\affiliation{Dipartimento di Fisica dell'Universit\`a di Roma
	``La Sapienza" and INFN, Sezione di Roma I, I-00185 Roma, Italy}

\author{Davide Rossini}
\affiliation{Dipartimento di Fisica dell'Universit\`a di Pisa
        and INFN, Largo Pontecorvo 3, I-56127 Pisa, Italy}

\author{Ettore Vicari} \altaffiliation{Authors are listed in
  alphabetic order.}  \affiliation{Dipartimento di Fisica
  dell'Universit\`a di Pisa, Largo Pontecorvo 3, I-56127 Pisa, Italy}

\date{\today}

\begin{abstract}
  We study the out-of-equilibrium dynamics of one-dimensional quantum
  Ising models in a transverse field $g$, driven by a time-dependent
  longitudinal field $h$ across their {\em magnetic} first-order
  quantum transition at $h=0$, for sufficiently small values of
  $|g|$. We consider nearest-neighbor Ising chains of size $L$ with
  periodic boundary conditions.  We focus on the out-of-equilibrium
  behavior arising from Kibble-Zurek protocols, in which $h$ is varied
  linearly in time with time scale $t_s$, i.e., $h(t)=t/t_s$.  The
  system starts from the ground state at $h_i\equiv h(t_i)<0$, where
  the longitudinal magnetization $M$ is negative.  Then it evolves
  unitarily up to positive values of $h(t)$, where $M(t)$ becomes
  eventually positive.  We identify several scaling regimes
  characterized by a nontrivial interplay between the size $L$ and the
  time scale $t_s$, which can be observed when the system is close to
  one of the many avoided level crossings that occur for $h\ge 0$. In
  the $L\to\infty$ limit, all these crossings approach $h=0^+$, making
  the study of the thermodynamic limit, defined as the limit
  $L\to\infty$ keeping $t$ and $t_s$ constant, problematic.  We study
  such limit numerically, by first determining the large-$L$ quantum
  evolution at fixed $t_s$, and then analyzing its behavior with
  increasing $t_s$.  Our analysis shows that the system switches from
  the initial state with $M<0$ to a positively magnetized state at $h
  = h_\star(t_s)>0$, where $h_\star(t_s)$ decreases with increasing
  $t_s$, apparently as $h_\star\sim 1/\ln t_s$.  This suggests the
  existence of a scaling behavior in terms of the rescaled time
  $\Omega = t \ln t_s/t_s$. The numerical results also show that the
  system converges to a nontrivial stationary state in the large-$t$
  limit, characterized by an energy significantly larger than that of
  the corresponding homogeneously magnetized ground state.
\end{abstract}

\maketitle


\section{Introduction}
\label{intro}

Out-of-equilibrium phenomena at first-order phase transitions have
been much investigated both in classical statistical models (see the
reviews~\cite{Binder-87, PV-24} and, e.g., Refs.~\cite{NN-75, FB-82,
  PF-83, FP-85, CLB-86, BK-90, LK-91, BK-92, VRSB-93, MM-00, LFGC-09,
  NIW-11, TB-12, ICA-14, PV-15, PV-16, PV-17, PV-17-dyn, LZ-17,
  PPV-18, Fontana-19, CCP-21, CCEMP-22}) and in quantum many-body
systems at zero temperature (see the reviews~\cite{Pfleiderer-05,
  RV-21,PV-24} and, e.g., Refs.~\cite{AC-09, YKS-10, JLSZ-10, LMSS-12,
  CNPV-14, CNPV-15, CPV-15, CPV-15-iswb, PRV-18, PRV-18-fowb, YCDS-18,
  RV-18, PRV-18-def, SW-18, LZW-19, PRV-20, DRV-20, SCD-21, TV-22,
  TS-23,Surace-etal-24}).  Since first-order transitions appear in
several different physical contexts, any progress in the theoretical
understanding of related nonequilibrium phenomena is of great
phenomenological importance.  The condensation of water, the melting
of ice, etc., are some examples of first-order classical transitions
at finite temperature.  First-order transitions driven by quantum
fluctuations occur in quantum Hall systems~\cite{PPBWJ-99}, itinerant
ferromagnets~\cite{VBKN-99, BKV-99}, heavy fermion
metals~\cite{UPH-04, Pfleiderer-05, KRLF-09}, SU($N$)
magnets~\cite{DK-16, DK-20}, quantum spin systems~\cite{LMSS-12,
  CNPV-14, LZW-19}, etc.  They display notable equilibrium and
out-of-equilibrium scaling behaviors, like classical and quantum
continuous transitions (see, e.g., Refs.~\cite{Fisher-74, Wilson-83,
  ZJ-book, PV-02, Sachdev-book, CPV-14, RV-21}).  In particular,
classical and quantum systems at first-order transitions turn out to
be particularly sensitive to the boundary conditions. Actually, the
sensitivity of the large-distance or low-energy properties to the
boundary conditions is one of the main distinctive differences between
the behaviors of finite-size systems at continuous and first-order
transitions~\cite{PV-24}.

In this paper we discuss the dynamics of one-dimensional quantum Ising
models in a transverse field, driven by a time-varying longitudinal
homogeneous field $h$ across the {\em magnetic} first-order quantum
transition (FOQT) at $h=0$. We consider periodic boundary conditions,
which preserve the ${\mathbb Z}_2$ symmetry of the model at
$h=0$. Moreover, they preserve translational invariance, which allows
us to focus on translation-invariant states, significantly simplifying
the numerical analysis.  We investigate the out-of-equilibrium
behavior arising from Kibble-Zurek (KZ) protocols, in which $h$ varies
linearly with time with time scale $t_s$, i.e., $h(t)=t/t_s$.  The
system starts from the negatively magnetized ground state at
$h_i\equiv h(t_i)<0$, and then evolves unitarily up to a positive
value $h(t)$, eventually leading to states with positive longitudinal
magnetization.

In the KZ dynamics, the time-dependent observables monitoring the
system develop an out-of-equilibrium finite-size scaling (OFSS)
behavior~\cite{PV-24, RV-21, PRV-18, PRV-18-fowb, RV-18, PRV-20,
  TV-22, TS-23} when $t_s\sim T(L)$, where $T(L)$ is the time scale to
make a transition from a magnetized state to the opposite one.  The
time scale $T(L)$ is related to the exponentially small gap $\Delta
\sim e^{-bL}$ at $h=0$, i.e., $T(L) \sim L / \Delta^2$. Thus,
$T(L)\sim e^{2bL}$ (apart from powers of $L$) increases very rapidly
with the system size.  When $t_s\gg T(L)$, the system evolves
adiabatically for $h \approx 0$, switching from the negatively
magnetized ground state to the positive one. On the other hand, for
$t_s\ll T(L)$, the passage through the $h=0$ avoided level crossing is
effectively instantaneous, so that the system persists in the {\em wrongly}
magnetized state ($M<0$) for $h(t) > 0$, as in the analogous
Landau-Zener problem for two-level systems~\cite{Landau-32, Zener-32,
  VG-96}. Therefore, the OFSS for $h\approx 0$ can be observed only
for relatively small systems or very large time scales
$t_s$~\cite{RV-21, PV-24}.

Beside the avoided level crossing at $h=0$, the low-energy spectrum of
finite-size systems shows a sequence of avoided level crossings
between the ${\rm wrongly}$ magnetized state and a discrete series of
zero-momentum kink-antikink states, labeled by $k=1,2, \ldots$. We can
associate a time scale $T_k(L)$ with each crossing, which is relevant
for the dynamics of the system when $t_s\ll T(L)$.  Since, the time
scales satisfy $T(L) \gg T_1(L) \gg T_2(L) \gg \ldots$ for
sufficiently large sizes, the time scale $t_s$ can be tuned in such a
way to select at which avoided crossing the system magnetization
changes sign.  More precisely, for $t_s \ll T(L)$, the negatively
magnetized state effectively survives across the $h=0$ and the first
$k-1$ avoided level crossings up to the one satisfying $t_s\approx
T_k(L)$. When $h(t) \approx h_k(L)$, the system jumps to a
kink-antikink state with positive magnetization.  A similar behavior
was also observed in Ref.~\cite{SCD-21}.

Since, the additional avoided crossings are localized at $h_k(L) = a/L
+ a_{1k}/L^{5/3} + O(L^{-2})$, they all collapse to $h=0^+$, for
$L\to\infty$. Thus, their role in the thermodynamic limit, defined as
the limit $L\to\infty$ keeping $t$ and $t_s$ constant, becomes unclear
and likely irrelevant.  We study such thermodynamic limit numerically,
by first determining the large-$L$ quantum evolution at fixed $t_s$,
and then analyzing the behavior with increasing $t_s$. As we shall
see, our results unveil the emergence of a peculiar infinite-size
scaling behavior.  We observe that the negatively magnetized state
jumps to states with positive magnetization at values $h_\star(t_s)>0$
that approach $h=0^+$ with increasing $t_s$, apparently as
$h_\star(t_s)\sim 1/\ln t_s$. This suggests an infinite-size scaling
behavior in terms of the scaling variable $\Omega = t \ln t_s/t_s$.
Another notable feature of the dynamics is that the quantum Ising
system approaches a nontrivial stationary state in the large-$t$
limit, characterized by an energy significantly larger than that of
the corresponding magnetized ground state. This difference is related
to the average work done to vary the field in the KZ protocol.

The paper is organized as follows. In Sec.~\ref{Isingchain} we present
the quantum Ising chain and discuss its low-energy spectrum along the
FOQT line. In particular, we discuss the small-$h$ behavior of the
energy of the lowest kink-antikink bound states, characterized by
nonanalytic $h^{2/3}$ corrections. We also identify a sequence of
avoided crossings between kink-antikink states and the {\em wrongly}
magnetized state, which plays a major role in the finite-size KZ
dynamics. In Sec.~\ref{equisca} we discuss the equilibrium finite-size
scaling close to such avoided level crossings. In Sec.~\ref{protocol}
we outline the KZ protocol, while in Sec.~\ref{dynsca} we analyze the
OFSS behavior for $h(t)\approx0$.  In Sec.~\ref{transfinite} we extend
the previous discussion to the avoided level crossings that occur for
$h > 0$.  The KZ dynamics in the infinite-size limit is numerically
investigated in Sec.~\ref{transLinfty}. In Sec.~\ref{conclu} we
summarize and draw our conclusions. Finally, App.~\ref{kkbound}
reports some analytical computations of the spectrum of the relevant
kink-antikink bound states at small transverse and longitudinal
external fields.

\section{First-order quantum transitions in quantum Ising chains}
\label{Isingchain}

\subsection{The model}
\label{model}

The nearest-neighbor quantum Ising chain in a transverse field is a
paradigmatic model showing continuous and FOQTs. Its Hamiltonian reads
\begin{equation}
  H_{\rm Is} = - J \, \sum_{\langle x,y\rangle} \sigma^{(1)}_x
  \sigma^{(1)}_{y}
  - g\, \sum_x \sigma^{(3)}_x - h \,\sum_x
  \sigma^{(1)}_x,
  \label{hedef}
\end{equation}
where $\sigma^{(k)}$ are the spin-$1/2$ Pauli matrices ($k=1,2,3$),
the first sum is over all nearest-neighbor bonds $\langle x,y\rangle$,
while the second and the third sums are over the $L$ sites of the chain.
The Hamiltonian parameters $g$ and $h$ represent homogeneous transverse
and longitudinal fields, respectively. Without loss of generality, we
assume $J = 1$, $g>0$. We also set the Planck constant $\hslash = 1$.

At zero temperature, and for $g=1$, $h=0$, the model~\eqref{hedef}
undergoes a continuous quantum transition belonging to the
two-dimensional Ising universality class, separating a disordered
phase ($g>1$) from an ordered ($g<1$) one. For any $g<1$, the
longitudinal field $h$ drives FOQTs along the $h=0$ line.

The low-energy properties at FOQTs crucially depend on the boundary
conditions, even in the $L\to\infty$ limit (see, e.g.,
Refs.~\cite{CNPV-14,CNPV-15, CPV-15,
  CPV-15-iswb,PRV-18-fowb,LMSS-12,RV-21,PV-24}).  We consider periodic
boundary conditions, preserving the ${\mathbb Z}_2$ symmetry and
translational invariance. The lowest energy levels for any $g<1$ and
$h=0$ are the magnetized states $|+\rangle$ and $|-\rangle$ along the
longitudinal direction, which spontaneously break the ${\mathbb Z}_2$
symmetry in the thermodynamic limit.  When varying $h$ across one
of the FOQT transition points ($h=0$, $g<1$), their (avoided) level
crossing gives rise to a discontinuity in the longitudinal magnetization
of the ground state $|\Psi_0\rangle$, 
\begin{equation}
  M= {1\over L} \langle \Psi_0 |\sum_x \sigma_x^{(1)} | \Psi_0 \rangle,
  \label{lomagn}
\end{equation}
in the thermodynamic limit~\cite{Pfeuty-70}:
\begin{equation}
  \lim_{h\to 0^\pm} \lim_{L\to\infty} M
  = \pm \, m_0, \qquad m_0 = (1 - g^2)^{1/8} .
  \label{sigmasingexp}
\end{equation}

\subsection{The spectrum}
\label{spectrum}

In a finite system of size $L$ with periodic boundary conditions,
quantum tunneling effects lift the degeneracy of the two
lowest magnetized states, which is present for $h=0$.
The Hamiltonian eigenstates are superpositions of the magnetized
$|\pm \rangle$ states,
\begin{equation}
  |0\rangle = {1\over \sqrt{2}} \left( |+\rangle + |-\rangle\right),
  \qquad 
  |1\rangle = {1\over \sqrt{2}} \left( |+\rangle - |-\rangle\right).
  \label{loweig}
\end{equation}
Their energy difference
\begin{equation}
  \Delta(L) \equiv E_1(L)-E_0(L)
  \label{deltadef}
\end{equation}
vanishes exponentially as $L$ increases. Indeed, its asymptotic
large-$L$ behavior is given by~\cite{Pfeuty-70,CJ-87}
\begin{equation}
  \Delta(L) \approx {2\over \sqrt{\pi L}} \, g^L.
  \label{deltapobc}
\end{equation}
The differences $E_n(L)-E_0(L)$ at the transition point $h=0$, between
the higher excited states ($n>1$) and the ground state, approach instead
finite values for $L\to \infty$.  We also recall that the size
dependence of the energy difference $\Delta(L)$ between the lowest
levels may drastically change when considering other boundary
conditions~\cite{CNPV-14,RV-21,PV-24}.  For example, in the case of
antiperiodic and opposite fixed boundary conditions, where the lowest
levels are single-kink configurations, the gap $\Delta(L)$ decays as 
$L^{-2}$ along the FOQT line, for any
$g<1$ (see, e.g., Refs.~\cite{CNPV-14, CPV-15-iswb}).

\begin{figure}[!t]
 \includegraphics[width=0.9\columnwidth, clip]{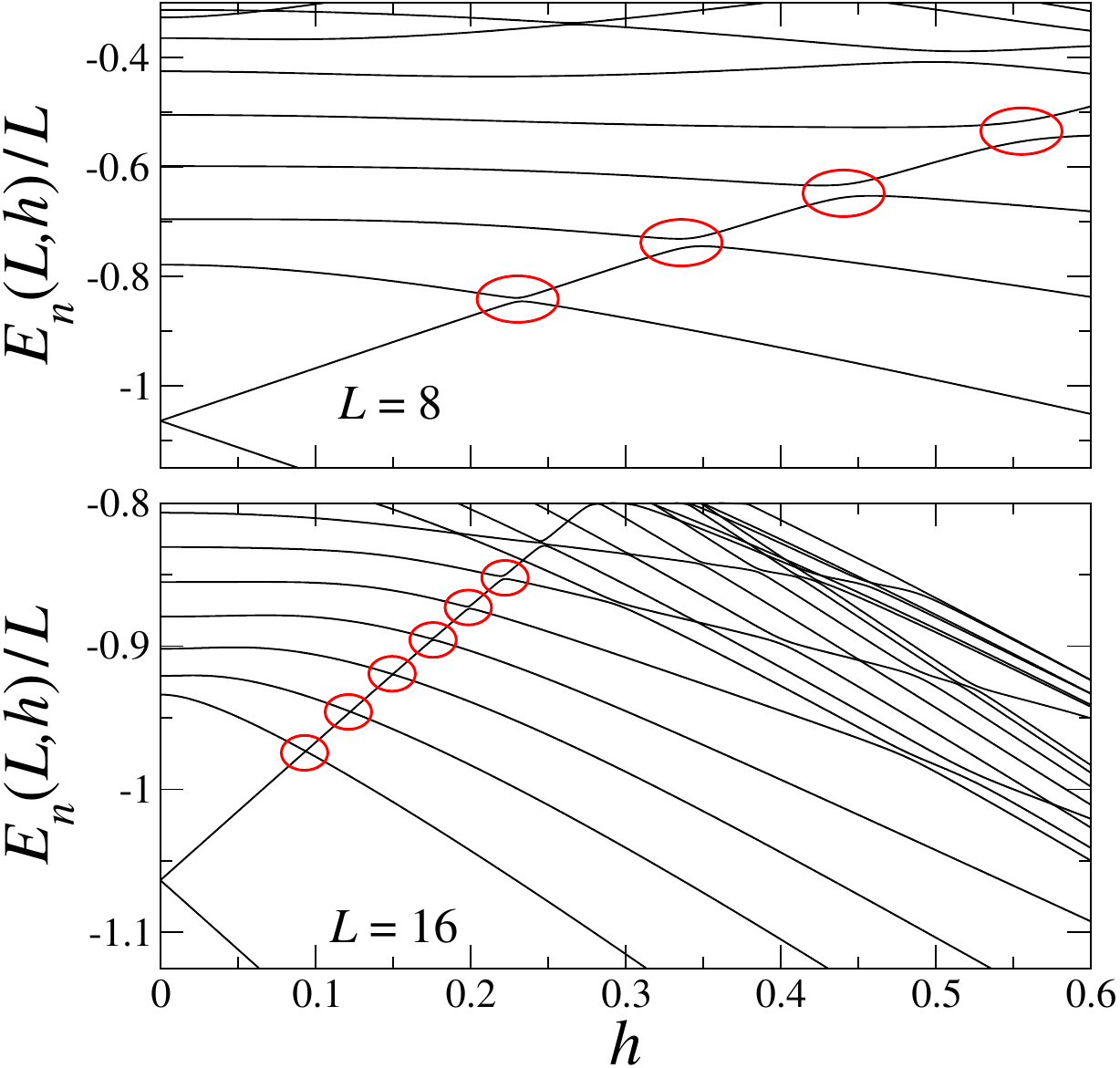}
 \caption{Spectrum of the quantum Ising chain~\eqref{hedef} with
   periodic boundary conditions at zero momentum, as a function of the
   longitudinal field $h$. Unless differently specified, here and in the following
   figures, we set $g=0.5$.  We plot the energy levels $E_n(L,h)$ as a
   function of $h$, for $L=8$ (top) and $L=16$ (bottom). Red circles
   denote the avoided crossings involving the magnetized state
   $|-\rangle$ and the lowest-energy zero-momentum kink-antikink states.
   Data are shown for $h\ge 0$, since the ${\mathbb Z}_2$ symmetry of
   the model implies that the spectrum is invariant under $h\to -h$.
   For the sake of clarity, in the bottom panel we report only the
   first 20 energy levels.}
 \label{zeromspec}
\end{figure}

As for the excitation spectrum, we first recall that periodic boundary
conditions preserve translational invariance, implying momentum
conservation even in finite systems.  This allows us to restrict our
analysis of the spectrum to the zero-momentum states, the only ones
that are relevant in dynamic processes preserving translational
invariance, i.e., when the initial condition and the external field driving
the dynamics are homogeneous.  Since the spectrum is symmetric with
respect to sign changes of the longitudinal field, i.e., under $h\to
-h$, it is sufficient to consider only nonnegative values ($h\ge 0$).

For sufficiently small values of $h$, the relevant low-energy states,
beside the magnetized $|\pm \rangle$ states, are zero-momentum
superpositions of states like
\begin{equation}
  |k_{w}\rangle = | \cdots  \uparrow_{x-1} \: \uparrow_x \:
  \downarrow_{x+1} \: \downarrow_{x+2} \cdots \downarrow_{x+w} \:
  \uparrow_{x+w+1} \cdots \rangle,
  \label{kakstates}
\end{equation}
which may be interpreted as kink-antikink bound states.  Results for
their energies in the presence of small longitudinal and transverse
fields can be found in
Refs.~\cite{MW-78,Rutkevich-08,Coldea-etal-10,Rutkevich-10}.  In
App.~\ref{kkbound} we reanalyze the problem, obtaining exact
finite-size results for their energy and magnetization.  These results
allow us to determine the pattern of the locations of the avoided
level crossings that characterize the low-energy spectrum of the
quantum Ising chain along the FOQT line.  Notice that the
classification of the low-energy Hamiltonian eigenstates in terms of
kink-antikink states is valid only if the magnetic energy $hL$ is
small enough, so that the states with no kinks, with one kink and one
antikink, with two kinks and two antikinks, etc, are separated in the
spectrum.  For $g=0$, the separation of these classes of states is $4
J$, so this requirement implies $h L \ll 4 J$, i.e., $h \ll 4/L$ for
$J=1$ and small values of $g$.

We supplement the above-mentioned analytic results with numerical
analyses of the spectrum of the quantum Ising chain (\ref{hedef}) for
$g=1/2$.  For this purpose, since the Ising-chain Hamiltonian is
nonintegrable for any $h \neq 0$, we use exact
diagonalization methods. Lanczos-based techniques allow us to compute
the exact representation of the first low-lying zero-momentum
eigenstates, for sizes up to $L \sim 26$ in the relevant Hilbert
subspace, which has a dimension of approximately $2.6 \times
10^6$~\cite{Sandvik-10}.

\begin{figure}[tbp]
 \includegraphics[width=0.9\columnwidth, clip]{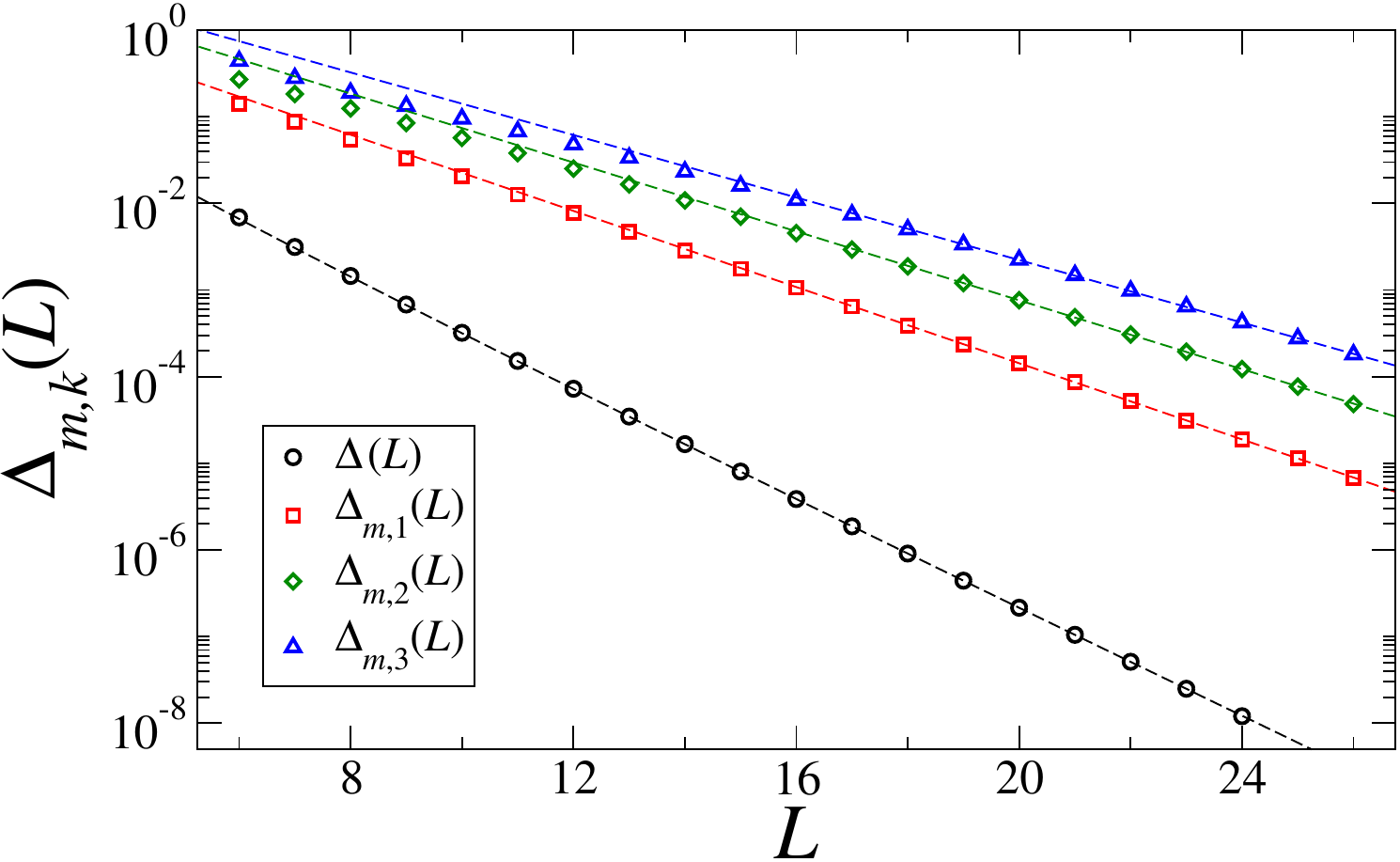}
 \caption{Exact-diagonalization results for the minimum energy
   differences $\Delta_{m,k}(L)$ between the magnetized state
   $|-\rangle$ and the first three zero-momentum kink-antikink states
   (colored data sets), as a function of the lattice size $L$.  We
   also show the energy gap $\Delta(L)$ of the lowest magnetized
   states at the integrable point $h=0$ (in this case we only
   considered sizes up to $L=24$, so that $\Delta(L) > 10^{-8}$,
   ensuring numerical convergence of our data). The dashed lines are
   the results of fits of the data for the largest chain sizes,
   typically for $L\ge 18$.  }
    \label{zeromgaps}
\end{figure}

In Fig.~\ref{zeromspec} we show the zero-momentum energy levels for
$g=1/2$, as a function of $h$ and for two different sizes $L$, as
obtained by exact diagonalization.  We note the presence of several
avoided crossings. The first one at $h=0$ has been discussed
above. The degeneracy of the magnetized states $|\pm\rangle$ at $h=0$,
is lifted for finite $L$, with an exponentially small gap, cf.
Eq.~\eqref{deltapobc}.  Other avoided level crossings occur for $h>0$
(red circles), involving the {\em wrongly} magnetized state
$|-\rangle$ and the kink-antikink states.  As it occurs for $h=0$, for
finite values of $L$ the degeneracy is lifted with an exponentially
small energy gap in the large-$L$ limit. To specify the location of
the avoided level crossings, we determine the size-dependent
longitudinal field $h_k(L)>0$ where the difference between the
energies of the $k$th kink-antikink state and of the magnetized
state $|-\rangle$ takes its minimum value $\Delta_{m,k}(L)$:
\begin{equation}
  \Delta_{m,k}(L) = \mathop{\rm min}_h \, [ E_k(L) - E_-(L) ] ,
  \quad k = 1,2, \ldots 
\end{equation}
As $L$ increases, this quantity behaves as $\Delta_{m,k}\sim
e^{-b_{k}L}$, see Fig.~\ref{zeromgaps}.  A numerical analysis of the
data shows that, for a fixed value of $L$, the gap $\Delta_{m,k}(L)$
increases with $k$, at least for the first few values of $k$;
moreover, it always satisfies $\Delta_{m,k}(L) > \Delta(L)$, where
$\Delta(L)$ is the gap between the lowest states at $h=0$.  Indeed, a
fit of our numerical results (the colored data sets in
Fig~\ref{zeromgaps}) for $18\le L \le 26$ to the Ansatz
\begin{equation}
  \ln \Delta_{m,k} = c_k - b_k\,L \,,
  \label{expans}
\end{equation}
gives $b_1 \approx 0.506$, $b_2 \approx 0.458$, $b_3 \approx
0.416$, which decrease with $k$ and are always smaller than
$b \approx 0.693$, the decay rate of the $h=0$ ground-state gap, 
obtained from Eq.~\eqref{deltapobc} with $g=1/2$.

The data for the zero-momentum spectrum in Fig.~\ref{zeromspec} also
show that the positions $h_k(L)$ of the avoided crossings approach $h=0$ 
approximately as $h_k(L)\sim 1/L$ with increasing $L$.
This statement can be justified more rigorously for small values of
$g$, by using the exact results (see App.~\ref{kkbound} and
Refs.~\cite{MW-78,Rutkevich-08,Coldea-etal-10,Rutkevich-10}) obtained
in the small-$g$ limit.  For sufficiently small values of $h$,
the energy of the magnetized state $|-\rangle$ can be written as
\begin{equation}
  E_{-}(L,h) \approx E_-(L,0) + m_0 \, h L  \,.
  \label{emagn}
\end{equation}
On the other hand, the energy of the zero-momentum kink-antikink
states is (see App.~\ref{AppA.2})
\begin{equation}
  E_k(L,h) = E_k(L,0) - m_0 \,h L + e_{k} \, g^{1/3} h^{2/3} +
  O(h),\;
  \label{ekink}
\end{equation}
where $k=1,2,\ldots$ labels the discrete levels, and the coefficient
$e_{k}>0$ increases with increasing $k$. As discussed in
App.~\ref{AppA.2}, the expansion~(\ref{ekink}) holds when $h$ is
small, but still satisfies $h \gg g/L^3$, and, in particular, in the
finite-size limit $h\to 0$, $L\to \infty$, at fixed (but not too
large, as discussed above) $h L$.  Subleading corrections (for fixed
values of $hL$) decay as $h^{m/3}$, with $m$ integer.  The coefficient
$e_{k}$ can be exactly computed: $e_{k} = 2 |\alpha_k|$, where
$\alpha_k < 0$ is the $k$th zero of the Airy function. Moreover, we
have $m_0 = 1$, as it also occurs for the energy of $|-\rangle$,
cf. Eq.~(\ref{emagn}). Therefore, the relevant kink states are fully
magnetized for large values of $L$, as discussed in App.~\ref{AppA.3}.

Equation~(\ref{ekink}) has been derived for small values of $g$.
We conjecture that a similar expansion holds for any value of $g < 1$
in the finite-size limit $h\to 0$, $L\to \infty$, at fixed $h L$.
Namely, we assume
\begin{eqnarray}
  E_k(L,h) &=& E_k(L,0) - m_0 h L \nonumber \\
  && + g^{1/3} h^{2/3} \Bigl( e_k + 
   \sum_{m=1}^\infty e_{k,m} \,h^{m/3}\Bigr), \quad
   \label{ekink2}
\end{eqnarray}
where $e_k$ and $e_{k,m}$ are functions of $g$ and $h L$, and the
prefactor $g^{1/3}$ has been added for consistency with
Eq.~(\ref{ekink}).  Eq.~(\ref{ekink2}) also predicts the behavior of
the magnetization of the kink states, since
\begin{equation}
  M_k(L,h) = - {1\over L} {\partial E_k \over \partial h}
  \approx m_0 - {2 e_{k}\over 3 L} \left( {g\over h}\right)^{1/3} \, .
  \label{magnkink}
\end{equation}
Note that the correction term is of order
$h^{-1/3} L = 1/(h L^3)^{1/3}$, which is small for $h L^3 \gg 1$, the
regime in which Eqs.~(\ref{ekink2}) and (\ref{magnkink}) apply.

The location $h_k(L)$ of the (avoided) level crossing between the
negatively magnetized state and the $k$th kink-antikink state
can be obtained by equating the energies
\begin{equation}
  E_-(L,h_k) = E_k(L,h_k).
  \label{crossing}
\end{equation}
Note that, for $h=0$,  the difference
\begin{equation}
  E_k(L,0)-E_-(L,0) = 4 + O(g) >0 
  \label{diffek0}
 \end{equation}
is finite in the large-$L$ limit and it is independent of $k$, since
the energy separation of the zero-momentum kink-antikink states
vanishes as $1/L^2$ for $h = 0$ [see Eq.~(\ref{Ekink-h0}) for $g$
small]. Therefore, at leading order, we obtain
\begin{equation} 
  h_{k}(L) = {a \over L}, \qquad
  a = \left. {E_k(L,0) - E_-(L,0)\over 2 m_0} \right|_{L\to\infty},
\label{defhk-a}
\end{equation}
where $a$ is independent of $k$. Thus, the avoided crossings occur for
values of the magnetic field $h$, where the expansion (\ref{ekink2})
holds.  Solving it perturbatively, we obtain the asymptotic expansion
\begin{equation}
  h_{k}(L) = {a \over L}  + \frac{a_{k}}{L^{5/3}} + \frac{a_{2k}}{L^2} +
  \frac{a_{3k}}{L^{7/3}} + \ldots 
\label{hkbeh}
\end{equation}
with $a_k = g^{1/3} e_{k} a^{2/3}/(2 m_0)$.
Eq.~(\ref{hkbeh}) implies that the difference of the locations of two
different level crossings is of order $L^{-5/3}$; more precisely, 
it behaves as
\begin{equation}
  h_{k'}(L) - h_{k}(L) = {a_{1,kk'} \over L^{5/3}} + {a_{2,kk'} \over L^2}
  + {a_{3,kk'}\over L^{7/3}} + \ldots 
  \label{hkdiff}
\end{equation}
Of course, this analysis neglects the interaction terms that cause the
crossings to be avoided for finite sizes.  However, the region where
the avoided level crossing takes place is exponentially small for
sufficiently large $L$, so Eq.~\eqref{hkbeh} can be applied to the
positions of the avoided crossings, defined by considering the
minimum of the energy difference between the two levels.

\begin{figure}[!t]
 \includegraphics[width=0.9\columnwidth, clip]{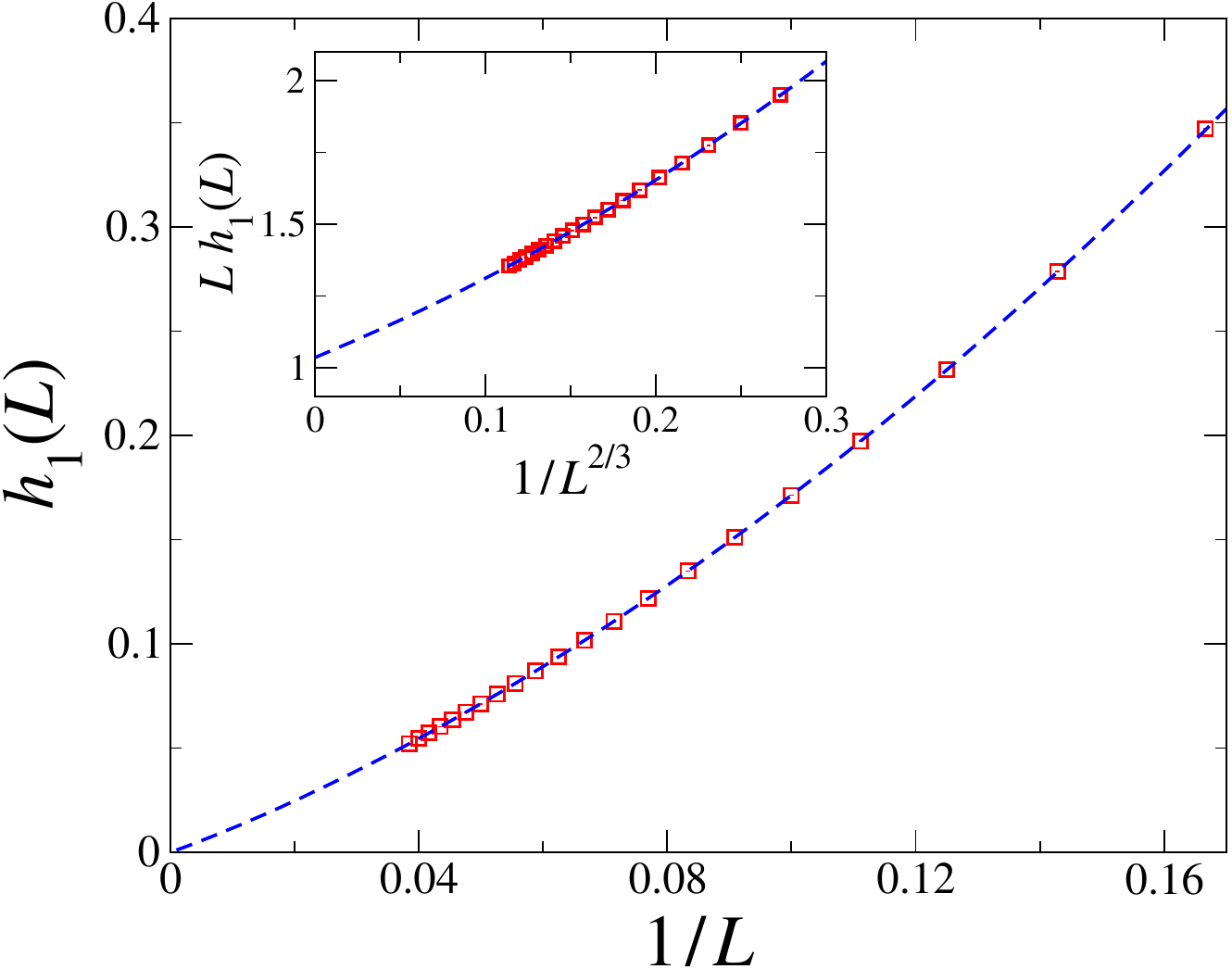}
 \caption{Location $h_1(L)$ of the level crossing between the
   negatively magnetized state and the lowest-energy kink-antikink
   state as a function of the inverse size $1/L$.  The absolute
   precision of the estimates of $h_1(L)$ is approximately $10^{-6}$
   for $L\le 11$, $10^{-7}$ for $12\le L \le 15$, and $10^{-9}$ for
   $16\le L \le 26$.  The errors are too small to be visible in the
   figure. The dashed line corresponds to the curve obtained by a
   four-parameter fit of the data for sizes $L\ge 19$ ($\chi^2/{\rm
     d.o.f}\approx 0.5$) to the expansion (\ref{hkbeh}), see also the
   text.  The inset shows a plot of $L\,h_1(L)$ versus $L^{-2/3}$, to
   highlight the predicted $L^{-2/3}$ subleading behavior, see
   Eq.~(\ref{hkbeh}).  }
    \label{loch1}
\end{figure}

The asymptotic large-$L$ predictions in Eqs.~(\ref{hkbeh}) and
(\ref{hkdiff}) are supported by the analysis of the numerical data
for $g=1/2$. In Fig.~\ref{loch1} we plot the available data for
$h_1(L)$, the value of $h$ where $E_1(L)-E_-(L)$ takes its minimum, up
to $L=26$.  Fits to the Ansatz (\ref{hkbeh}) with four free parameters
including only data with $L\ge L_{\rm min}$ give results which are barely
dependent on $L_{\rm min}$.  In particular, for $L_{\rm min} = 19$ (we
include the 8 largest sizes) we obtain $a= 1.0352(1)$, $a_1=2.521(4)$,
$a_{21}=- 0.50(1)$, and $a_{31}=3.97(2)$ with $\chi^2/{\rm d.o.f}\approx 0.5$.
The corresponding curve is displayed in dashed blue color.

As discussed in Ref.~\cite{PRV-18-fowb}, in the case of fixed parallel
boundary conditions the location of the avoided crossing between the
magnetized and lowest-energy kink-antikink state also has an expansion
as reported in Eq.~(\ref{hkbeh}). Moreover, as discussed in
App.~\ref{AppA.4}, the coefficient $a$ of the $1/L$ term should be the
same as for periodic boundary conditions.  This is confirmed by our
results.  The estimate $a\approx 1.0352$ obtained here for periodic
boundary conditions is substantially consistent with the result
$a=1.0370(5)$ reported in Ref.~\cite{PRV-18-fowb} for fixed parallel
boundary conditions.  We may also compare the estimate of $a_1$ with
the small-$g$ approximation $a_1 \approx e_{1} g^{1/3} a^{2/3}/(2m_0)$,
where $e_{1}=2|\alpha_1|$ ($\alpha_1$ is the smallest
zero of the Airy function, $|\alpha_1|\approx 2.338$), see
Eq.~(\ref{Enapprox}).  We obtain $a_1\approx 1.97$ for $g=1/2$ (using
$a\approx 1.035$ and $m_0\approx 0.96$), which is reasonably close to
the actual estimate $a_1\approx 2.52$ obtained by the fit of our data
for $g=1/2$.

\begin{figure}[!t]
 \includegraphics[width=0.9\columnwidth, clip]{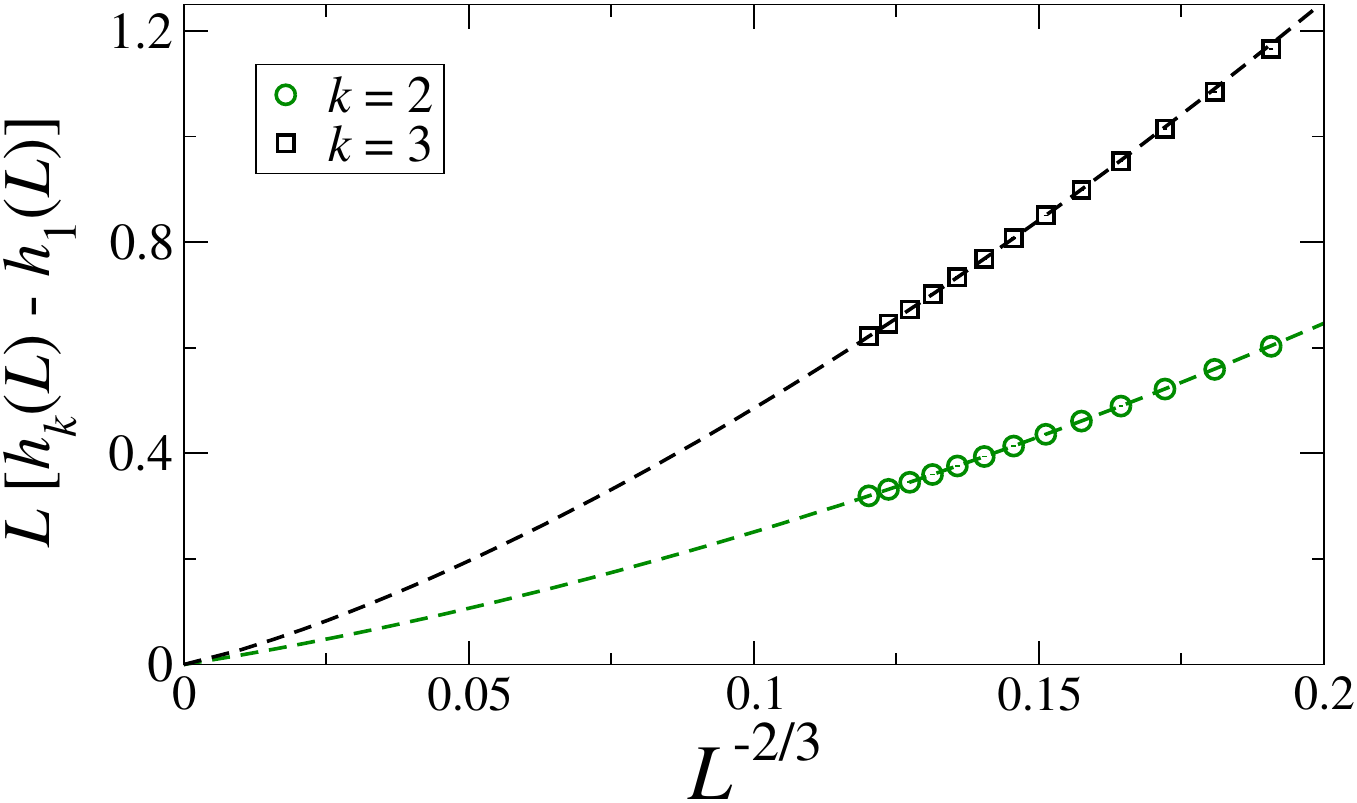}
 \caption{
   Plot of $L[h_k(L)-h_1(L)]$ versus $L^{-2/3}$, to highlight the
   expected behavior $h_k(L)-h_1(L) \sim L^{-5/3}$.  The precision of
   the estimates of $h_k(L)$ with $k=2,3$ is approximately $10^{-6}$ 
   for $L\le 11$ and $10^{-7}$ for $L\ge 12$.}
    \label{locdiff}
\end{figure}

Finally, in Fig.~\ref{locdiff} we show the differences
$h_k(L)-h_1(L)$ for $k=2$ and $k=3$, up to $L=24$. Their behavior is
consistent with the asymptotic formula~(\ref{hkdiff}).
Three-parameter fits to that expansion give
$a_{1,12}=1.65(1)$ (using data for $19 \leq L \leq 24$, with
$\chi^2/{\rm d.o.f.}\approx 0.6$) and $a_{1,13}=1.98(4)$ (using data
for $L\ge 21$, with $\chi^2/{\rm d.o.f.}\approx 0.6$).

\section{Equilibrium finite-size scaling at the avoided level crossings}  
\label{equisca}

At FOQTs the low-energy properties satisfy general equilibrium
finite-size scaling (EFSS) laws as a function of the field $h$ and of
the system size $L$~\cite{CNPV-14,PRV-18-fowb,CNPV-15,PV-24}.  In the
EFSS framework, when the boundary conditions preserve the $h\to -h$
symmetry, the relevant scaling variable is the ratio~\cite{CNPV-14}
\begin{equation}
  \Phi = {\delta E(L,h) \over \Delta(L)} =  { 2 m_0 h L \over \Delta(L)},
  \label{phidef}
\end{equation}
where $\delta E(L,h)=E(L,h) - E(L,-h)= 2 m_0 hL$ quantifies the magnetic energy
due to addition of the longitudinal field $h$, while
$\Delta(L) \equiv \Delta(L,h=0)$ in the denominator is the ground-state
energy gap at $h=0$.
The zero-temperature EFSS limit corresponds to $L \to \infty$ and $h
\to 0$, keeping $\Phi$ fixed.  In this limit, the ground-state magnetization
$M(L,h)$ and the energy difference $\Delta(L,h)$ of the lowest levels
asymptotically behave as~\cite{CNPV-14}
\begin{equation}
  M(L,h) \approx m_0 \, {\cal M}(\Phi) \,,
  \quad
  \Delta(L,h) \approx \Delta(L) \, {\cal D}(\Phi) \,.
  \label{efssm}
\end{equation}
An analogous EFSS behavior is expected for other observables, such as
the ground-state fidelity~\cite{RV-18}, and at finite 
temperature~\cite{RV-21}.

Unlike continuous quantum transitions, the EFSS at FOQTs drastically
depends on the nature of the boundary
conditions~\cite{CNPV-14,PRV-18,PRV-18-fowb,RV-21,PV-24}.  In
particular, this is evident when the scaling variable $\Phi$ is
expressed in terms of $L$ and $h$, given that the size behavior of
$\Delta(L)$ at the FOQT crucially depends on the boundary
conditions. Indeed, as already mentioned, the gap $\Delta$ may have
either an exponential or a power dependence on $L$, depending on the
boundary conditions, although in all cases the finite-size structure
of the low-energy spectrum must lead to the discontinuous
equilibrium behavior characterizing the FOQTs in the thermodynamic
limit.  On the other hand, in continuous quantum transitions the
critical power behavior cannot be changed by the boundary conditions.

With neutral boundary conditions, such as periodic boundary
conditions, the magnetized states $|\pm\rangle$ represent the
lowest-energy excitations.  As discussed in Sec.~\ref{spectrum},
$\Delta(L) \sim e^{-b L}$ at the FOQT point, while the energy
differences $\Delta_n\equiv E_n-E_0$ associated with the higher
excited states ($n>1$) are finite (more generally, $\Delta/\Delta_n$
decreases exponentially, with possible power corrections) in the
large-$L$ limit. For sufficiently large $L$ and $h\ll 1$, the
low-energy properties close to the avoided level crossing can be
obtained by restricting the theory to the two lowest-energy states
$|0\rangle$ and $|1\rangle$, or equivalently to the magnetized states
$| + \rangle$ and $| - \rangle$~\cite{CNPV-14,PRV-18}. In this
restricted Hilbert space, the lowest two-level spectrum can be
effectively described by a two-level Hamiltonian
\begin{equation}
  H_{2 {\rm lev}} = \varepsilon \, \sigma^{(3)} + \zeta \, \sigma^{(1)},
  \label{hrtds}
\end{equation}
where the parameters correspond to $\varepsilon=m_0 h L$ and
$\zeta=\Delta(L)/2$. This effective two-level reduction allows us to
exactly compute the EFSS functions of the magnetization $M(L,h)$ and
gap $\Delta(L,h)$~\cite{CNPV-14}.  The convergence to the asymptotic
two-level EFSS behavior is generally fast, being controlled by the
ratio between the exponentially suppressed gap $\Delta$ and the
energy-level differences with the higher states, which are finite for
$L\to\infty$.

We remark that the above two-level EFSS behavior arises because only
two states, the magnetized states $|\pm\rangle$, are degenerate in the
infinite-volume limit at $h=0$. A different behavior emerges in other
cases, as for antiperiodic boundary conditions. Indeed, in that
situation, an infinite number of states (the single-kink states)
become degenerate in the infinite-volume limit.  The presence of this
infinite tower of degenerate states changes the size behavior of
$\Delta(L)$, which scales as $1/L^2$ and not exponentially in $L$. 
Thus, the scaling behavior (\ref{efssm}) in terms of the variable
$\Phi$ defined in Eq.~(\ref{phidef}) still holds, but a two-level
description of the scaling behavior is no longer valid (see, e.g.,
Refs.~\cite{CNPV-14,PRV-18-fowb,RV-21,PV-24}).

The two-level truncation and the corresponding scaling behavior relies
on a single basic assumption, that the energy difference $\Delta(L)$
between the two considered states vanishes in the large-$L$ limit
faster than the energy differences $\Delta_n(L)$ with the neglected
states, i.e., $\Delta(L)/\Delta_n(L)\to 0$ as $L\to \infty$.  Under
this condition, two-level scaling holds, provided the variable $\Phi$
(or the equivalent variable $\varepsilon$) vanishes at the avoided
crossing. Thus, the numerator in the definition of $\Phi$ should be
equal to $2m_0 (h -h_{\rm cr}) L$, where $h_{\rm cr}$ is the value of
the magnetic field at the avoided crossing. Such a scaling behavior
was indeed verified for localized magnetic fields~\cite{CNPV-14} and
for fixed boundary conditions \cite{PRV-18-fowb}.

\begin{figure}[!t]
 \includegraphics[width=0.9\columnwidth, clip]{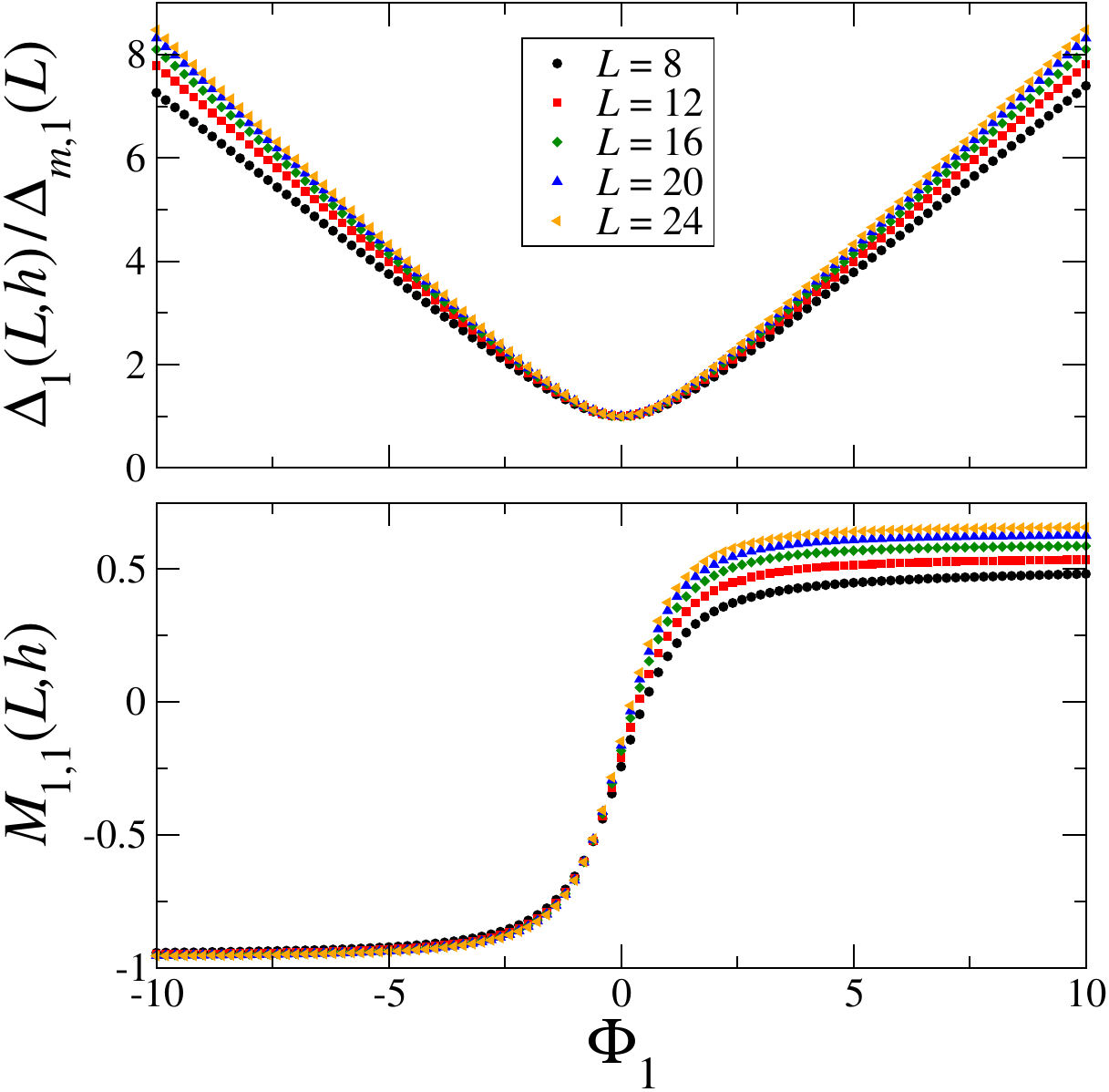}
 \caption{The gap $\Delta_1$ (top) and the magnetization
   $M_{1,1}$ (bottom) close to the avoided crossing of the negatively
   magnetized state and of the lowest-energy kink-antikink state
   occurring at $h\approx a/L + a_1/L^{5/3}$, with $a\approx 1.035$
   and $a_1=2.52$. We report the data for different values of the
   longitudinal field, as a function of the scaling variable $\Phi_1$
   defined in Eq.~\eqref{phi2def}.  We consider the magnetization of the
   state that has the lowest energy.  The various data sets refer to
   different chain lengths, from $L=8$ to $24$ (see legend). These
   results can be also interpreted as the behavior at the
   pseudocritical transition $h = h_1(L)$ for the reduced model
   (\ref{redisH}). }
    \label{m1fss}
\end{figure}

The previous conditions are satisfied at all avoided crossings
discussed in Sec.~\ref{spectrum}, thus we expect a two-level
scaling behavior also in those cases.  For $h$ close to $h_k(L)$,
we expect a scaling behavior in terms of
\begin{equation}
  \Phi_{k} = {2m_0[h-h_{k}(L)] L\over \Delta_{m,k}(L)},
  \label{phi2def}
\end{equation}
where $h_k(L)$ is the position of the avoided crossing  and 
$\Delta_{m,k}(L)$ is the corresponding gap.
The magnetization of the two levels 
asymptotically behaves as
\begin{equation}
  M_{k,a}(L,h) \approx {\cal M}_{a}(\Phi_k),
\end{equation}
where $a=1,2$ labels the two states present at the crossing, while the
energy gap scales as
\begin{equation}
  \Delta_{k}(L,h) \approx \Delta_{m,k}(L) \, {\cal E}(\Phi_k).
\end{equation}
These EFSS behaviors are nicely supported by our numerical results
shown in Fig.~\ref{m1fss} for $k=1$.  The scaling functions can be
computed using the effective two-level model, as discussed in
Ref.~\cite{PRV-18-fowb}.

The behavior mentioned above also emerges if one considers a 
reduced Hamiltonian in which one of the two magnetized states is 
projected out, i.e., if one considers the reduced Hamiltonian
\begin{equation}
  H_{{\rm Is}-0} = [1-\Pi_0(h)]\,H_{\rm Is} [1-\Pi_0(h)],
  \label{redisH}
  \end{equation}
where $\Pi_0(h)$ is, for each $h$, the projector on the ground state
of the system. This model also undergoes a FOQT. Indeed, if we take
the infinite-size limit at fixed $h > 0$, we obtain $M = m_0$ for the
magnetization: As soon as $ L > a/h$ [$a$ is the constant defined in
  Eq.~(\ref{defhk-a})], the ground state is the positively magnetized
kink-antikink state. Analogously, by symmetry, $M = - m_0$ for $h<
0$. Thus, the magnetization is discontinuous for $h=0$, signaling the
FOQT. However, the discontinuity is not related to a closing gap for
$h=0$, as the gap between the lowest states remains finite in the
$L\to\infty$ limit at $h=0$.  A vanishing gap can be observed,
however, by taking a less conventional infinite-size limit, i.e., if we
take $L\to\infty$ and $h =0$ simultaneously, keeping $hL$ fixed. In
this case for $hL = a$ we have a pseudotransition, with an
exponentially small ground-state gap and a discontinuous ground-state
magnetization.  For $hL > a$ we have $M=m_0$, as the kink state is the
ground state of the model, while, for $hL < a$, we have $M = - m_0$,
as the ground state of the model is the $|-\rangle$ state. Note that
the same type of behavior can be observed changing the boundary
conditions.  If we fix the boundary spins to $-1$---it represents an
equivalent method to project out the positively magnetized state---we
obtain the same behavior \cite{PRV-18-fowb}.

As a final comment, note that the FOQT is quite robust and indeed, it
would survive even after projecting out both magnetized states: we would
obtain a behavior similar to that observed for antiperiodic boundary
conditions. In this case, the kink-antikink bound states represent the
lowest-energy states of the model.

\section{Dynamic protocol across first-order quantum transitions}
\label{protocol}
  
To investigate the out-of-equilibrium behavior that arises when
crossing the FOQTs of the quantum Ising chain, we focus on a dynamic
protocol in which the longitudinal field varies across the
value $h=0$, for $g<1$.  We consider the simplest linear time dependence
\begin{equation}
  h(t) = t/t_s,
  \label{hst}
\end{equation}
where $t_s$ is the corresponding time scale.  For $t=0$, the
longitudinal field vanishes and the system goes across the FOQT.
The evolution starts at time $t_i=h_i t_s<0$ (we assume $h_i < 0$, so
$h(t_i) = h_i<0$) from the corresponding ground state
$|\Psi(t=t_i)\rangle \equiv |\Psi_0(h_i)\rangle$, with negative
magnetization
\begin{equation}
M_i =\langle \Psi_0(h_i) | \sigma_x^{(1)} |
\Psi_0(h_i)\rangle \lesssim -m_0.
\label{miniti}
\end{equation}
If $|h_i|$ is sufficiently small, then $M_i \approx - m_0$.  For
$t>t_i$, the field varies according to Eq.~\eqref{hst} and the system
evolves unitarily according to the Schr\"odinger equation
\begin{equation}
  i{{\rm d} \, |\Psi(t)\rangle \over {\rm d} t} =
  H[h(t)] \, |\Psi(t)\rangle \,, 
  \label{unitdyn}
\end{equation}
up to a time $t=t_f>0$, corresponding to $h(t_f)=h_f>0$, which is
sufficiently large to obtain states $|\Psi(t)\rangle$ with positive
longitudinal magnetization.

This protocol resembles the one considered for the study of the
so-called KZ problem, i.e., of the scaling behavior of the amount of
defects when a system slowly moves across a continuous quantum
transition~\cite{Kibble-76, Kibble-80,Zurek-85, Zurek-96, ZDZ-05,
  PG-08, PSSV-11, CEGS-12, RV-21}.  An analogous KZ protocol for
quantum Ising chains at their FOQTs was also considered in
Ref.~\cite{SCD-21} addressing nucleation phenomena, and in
Ref.~\cite{Surace-etal-24} focusing on string-breaking phenomena.

The time behavior of the system for $t>t_i$ can be monitored by
computing the instantaneous longitudinal magnetization
\begin{equation}
  M(t) = {1\over L}\sum_{x=1}^L \langle\Psi(t)|
  \sigma_{x}^{(1)}|\Psi(t) \rangle.
  \label{mxm}
\end{equation}
In particular, with periodic boundary conditions, $M(t) =
\langle\Psi(t)| \sigma_{x}^{(1)}|\Psi(t) \rangle$, due to
translational invariance.  We also consider the excess energy, defined
as the difference between the time-dependent energy density and the
energy density of the ground state at the instantaneous field $h(t)$,
\begin{eqnarray}
  E_{\rm ex}(t) & = & {1\over L}
  \langle\Psi(t)\,|\,H_{\rm Is}[h(t)]\,|\Psi(t) \rangle \nonumber \\
  & & - {1\over L}
  \langle\Psi_0[h(t)]\,|\,H_{\rm Is}[h(t)]\,| \Psi_0[h(t)] \rangle. \qquad
  \label{eneexdef}
\end{eqnarray}
We also define an alternative quantity, considering only the hopping term
of the Hamiltonian~\eqref{hedef}, 
\begin{eqnarray}
  K_{\rm ex}(t) & = & - {1\over L}
  \langle\Psi(t)\,|\sum_{\langle x,y\rangle} \sigma^{(1)}_x
  \sigma^{(1)}_{y} \,|\Psi(t) \rangle \nonumber \\
  & & + {1\over L} 
  \langle\Psi_0[h(t)]\,|\sum_{\langle x,y\rangle} \sigma^{(1)}_x
  \sigma^{(1)}_y\,| \Psi_0[h(t)] \rangle. \qquad
  \label{eneexdef2}
\end{eqnarray}
Of course, $E_{\rm ex}(t=t_i) = K_{\rm ex}(t=t_i)=0$.  The 
differences $E_{\rm ex}(t)$ and $K_{\rm ex}(t)$ somehow quantify the
degree of nonadiabaticity of the evolution.
Another quantity that provides useful information on the evolving
state is the adiabaticity function, defined as
\begin{equation}
  A(t)  = |\langle \Psi_0[h(t)] \, | \, \Psi(t) \rangle | ,
  \label{adiabfunc}
\end{equation}  
see, e.g., Ref.~\cite{TV-22}.

We have implemented the KZ protocol numerically.  The Schr\"odinger
equation~\eqref{unitdyn} is integrated by means of a standard
fourth-order Runge-Kutta approach. We choose a sufficiently small
time step $dt = 2.5 \times 10^{-3}$, to ensure convergence of all
our results up to the largest considered sizes ($L=20$)
and times ($t_s = 10^6$).

\section{Out-of-equilibrium finite-size scaling}
\label{dynsca}

To study the out-of-equilibrium behavior arising in the KZ
protocol outlined in Sec.~\ref{protocol}, we identify dynamic
scaling regimes for large $L$ and $t_s$, essentially related to the
avoided level crossings of the spectrum.  In this section we adopt the
OFSS framework developed for FOQTs (see, e.g.,
Refs.~\cite{PRV-18,PRV-18-def,PRV-20,RV-21,PV-24}).

\subsection{Out-of-equilibrium finite-size scaling at the $h=0$
avoided crossing}
\label{h0outfss}

In the simultaneous limits $t_s\to\infty$ and $L\to\infty$, the system
undergoes an OFSS behavior for $h=0$, due to the avoided crossing of
the two magnetized states.  Since the EFSS behavior should be
recovered when $t_s$ is much larger than the time scale of the
low-energy modes, one scaling variable should be identified with the
corresponding equilibrium scaling variable.  Replacing $h$ with $h(t)$
in the definition~\eqref{phidef} of $\Phi$, we obtain
\begin{equation}
  \widehat \Phi \equiv {2 m_0 h(t) L \over \Delta(L)} =
           {2 m_0 t L \over t_s\Delta(L)}.
  \label{katdef}
\end{equation}
The second scaling variable can be naturally defined as
\begin{equation}
  \Theta \equiv t\,\Delta(L) .
  \label{thetadeffo}
  \end{equation}
It is useful to define an additional scaling variable that does not
depend on the time $t$, which can be interpreted as the ratio between
$t_s$ and the time scale $T(L)$ that characterizes the crossing of the
transition point $h=0$ for a system of size $L$:
\begin{equation}
  \Upsilon = {\Theta \over \widehat\Phi} = {t_s \over T(L)},\qquad
  T(L)={2m_0 L\over \Delta(L)^2} \,.
  \label{upsilondef}
\end{equation}
The OFSS limit corresponds to $t, t_s, L\,\to\infty$, keeping
$\widehat\Phi$ and $\Upsilon$ fixed. In this limit, the magnetization
is expected to scale as~\cite{PRV-18-def,PRV-20,RV-21,PV-24}
\begin{equation}
  M(t,t_s,h_i,L) \approx m_0\,{\cal M}(\Upsilon, \widehat\Phi) \,,
  \label{mtsl}
\end{equation}
independently of $h_i<0$.  The adiabatic limit corresponds to
$t,t_s\to \infty$ at fixed $L$ and $t/t_s$, i.e., to $\Upsilon\to \infty$.
In this limit, ${\cal M}(\Upsilon, \widehat\Phi)$ should
reproduce the EFSS behavior~\eqref{efssm} with $\widehat\Phi=\Phi$.
The scaling functions are expected to be universal (i.e., independent
of $g$ along the FOQT line, for a given class of boundary conditions).

Note that the OFSS occurs in a narrow interval of longitudinal fields.
Indeed, since $\widehat{\Phi}$ is kept fixed in the OFSS limit, the
relevant scaling behavior develops in the interval
\begin{equation}
  |h(t)| \lesssim L^{-1} \Delta(L),
  \label{hrange}
\end{equation}
which rapidly shrinks with increasing $L$, as $|h|\sim e^{-bL}$, apart
from powers of $L$.  This implies that the OFSS behavior must be
independent of the initial value $h_i<0$, when it is kept fixed in the
large-$L$ limit.

The OFSS functions can be computed by
exploiting a two-level effective theory.  Indeed, since the low-energy
behavior is controlled by the two lowest-energy states, we can again
consider the effective Hamiltonian~(\ref{hrtds}), which now becomes
time-dependent:
\begin{align}
  H_{r} = & \:\: \varepsilon(t) \, \sigma^{(3)} + \zeta \,
  \sigma^{(1)} ,
  \nonumber \\
  \varepsilon(t) = & \:\: {m_0 \, h(t) \, L}
  = {m_0 \, t L \over t_s} \,, \quad \zeta = {\Delta(L)\over 2} .
  \label{hrdef2t}
\end{align}
The system is thus equivalent to a two-level quantum system in which
the energy separation of the two levels is a linear function of time,
with the correspondence
\begin{equation}
  \widehat{\Phi} = {\varepsilon(t)\over \zeta}, \qquad
  \Upsilon = {2 \zeta^2 t\over \varepsilon(t)}.
\end{equation}
This dynamics was first investigated by Landau and
Zener~\cite{Landau-32, Zener-32} and then solved exactly in
Ref.~\cite{VG-96}.  The OFSS function ${\cal
  M}(\Upsilon,\widehat\Phi)$ defined in Eq.~\eqref{mtsl} can be
computed by taking the expectation value of $\sigma^{(3)}$ over the
solution $|\Psi_{2l}(t)\rangle$ of the Schr\"odinger equation with the
initial condition $|\Psi_{2l}(t_i)\rangle = |-\rangle$ (where
$|-\rangle$ is the eigenstate of $\sigma^{(3)}$ with negative
eigenvalue), i.e.~\cite{PRV-18-def, PRV-20}
\begin{eqnarray}
{\cal M}(\Upsilon, \widehat\Phi) & = & \langle \Psi_{2l}(t)
  | \sigma^{(3)}|\Psi_{2l}(t)\rangle    \label{fsigmasol}\\
  & = & -1 + \tfrac12 \Upsilon e^{-{\pi
      \Upsilon\over 8}} \left| D_{-1+i{\Upsilon\over 4}} (e^{i{3\pi\over
      4}} \widehat{\Phi}\,\Upsilon^{1\over 2}) \right|^2,
\nonumber
\end{eqnarray}
where $D_\nu(x)$ is the parabolic cylinder function~\cite{AS-1964}.
In the large-time limit, or, equivalently, for
$\widehat{\Phi}\to\infty$ keeping $\Upsilon$ fixed, we obtain
\begin{equation}
  {\cal M}_{t\to\infty}(\Upsilon) = 
  {\cal M}(\Upsilon, \widehat\Phi\to\infty) =
  1 - 2 \, e^{-{\pi \Upsilon\over 2}}.
\label{asyLZfo}  
\end{equation}
Therefore, the large-time magnetization remains negative,
${\cal M}_{t\to\infty}(\Upsilon)<0$, for
$\Upsilon < {2\ln 2/\pi}$.
In particular, ${\cal M}_{t\to\infty}=-1$ for $\Upsilon\to 0$, i.e.,
for $t_s\ll T(L)$. In this case, the system is still negatively
magnetized as at the beginning of the dynamics, because the passage
across $h=0$ is effectively instantaneous.  Instead, for large values
of $\Upsilon$, corresponding to $t_s\gg T(L)$, the magnetization is
positive, ${\cal M}_{t\to\infty}(\Upsilon) >0$ and, in particular,
${\cal M}_{t\to\infty}(\Upsilon) \to 1$ for $\Upsilon\to\infty$.

It is important to note that the {\em thermodynamic} limit may
formally be obtained by taking the limit $\Upsilon\to 0$ in the
previous equations, keeping the $L$-independent scaling variable
\begin{equation}
  W = {t^2 \ln |t|\over t_s}
  \label{Wdef}
\end{equation}
fixed ($W$ can be derived by appropriately combining $\widehat{\Phi}$
and $\Theta$).  However, the scaling behavior~\eqref{fsigmasol} becomes
trivial in this limit~\cite{PRV-18-def}, reflecting the fact that the
limit $\Upsilon\to 0$ corresponds to small values of $t_s$, which do
not allow the system to make a transition to the positive magnetized
state.  Moreover, the interval of values of $h$ in which OFSS applies
shrinks to zero.  As we shall see below, in the infinite-volume limit
a more complex behavior emerges, involving successive avoided level
crossings with a peculiar, apparently independent, scaling behavior.

\subsection{Out-of-equilibrium finite-size scaling at the avoided
  level crossings for $h>0$}
\label{hl0outfss}

Let us now investigate whether it is possible to identify additional
out-of-equilibrium scaling regimes associated with the other avoided
crossings that occur for $h=h_k(L)>0$. We show here that one can
observe further nontrivial OFSS behaviors whenever the time scale
$t_s$ is significantly smaller that the time scale $T(L)$, defined in
Eq.~(\ref{upsilondef}), which controls the dynamics at $h=0$.  Indeed,
for $t_s \ll T(L)$, corresponding to $\Upsilon\ll 1$, the system does
not jump to the positively magnetized state when crossing the
transition $h=0$. The dynamics is so fast that the passage can be
considered as effectively instantaneous.  The question is whether the
system, which is still negatively magnetized, can then make a
transition to the positively magnetized lowest-energy kink state at
$h\approx h_{1}(L)$, with a corresponding OFSS regime.

To define an OFSS regime for $h(t)\approx h_1(L)$, we simply
generalize the definitions given in Sec.~\ref{h0outfss}, as already
done in Ref.~\cite{PRV-18-fowb} for fixed parallel boundary
conditions.  We define
\begin{equation}
  \hat{t} \equiv t + h_{1}(L)\,t_s,
  \label{trdef}
\end{equation}
such that $\hat{t} = 0$ corresponds to the pseudo-transition point at
$h = h_{1}(L)$.  The natural scaling variables are
\begin{equation}
 \widehat\Phi_{1} = {2 m_0 L \over \Delta_{m,1}(L)} {\hat{t} \over t_s},
 \qquad
\Theta_{1} \equiv \hat{t}\,\Delta_{m,1}(L),
 \label{2cscavardef}
\end{equation} 
and 
\begin{equation}
  \Upsilon_{1} = {\Theta_{1}\over \widehat\Phi_{1}} =
          { t_s \over T_{1}(L)}, \qquad
          T_{1}(L) = {2m_0L\over \Delta_{m,1}(L)^2}.
  \label{upsdef1}
\end{equation}
OFSS is obtained by taking $\hat{t},t_s,L\,\to\infty$, keeping the
scaling variables $\Upsilon_{1}$ and $\widehat{\Phi}_{1}$ fixed. In this
limit, the magnetization obeys the asymptotic OFSS behavior
\begin{equation}
  M(L,t_s,t) \approx {\cal M}_{1}(\Upsilon_{1}, \widehat\Phi_{1}).
  \label{mtsl1}
\end{equation}
The OFSS functions can again be computed using an effective two-level
model, but now we should take into account that the magnetization of
the two relevant states can be different. This issue is discussed in
Ref.~\cite{PRV-18-fowb}. In particular, if the magnetization of the
kink-antikink state is $m_k$ and $r_k=m_k/m_0$, Eq.~(\ref{asyLZfo})
is replaced by
\begin{equation}
{\cal M}(\Upsilon_1,\widehat{\Phi}_1\to\infty) = r_k -
(1 + r_k) e^{-{\pi \over 1 + r_k} \Upsilon}.
\label{Masymptotico_KZ}
\end{equation}
Since $\Delta_{m,1}(L)/\Delta(L) \sim e^{(b -
  b_1)L} \gg 1$ (see Sec.\ref{spectrum}), the ratio
\begin{equation}
  T_1(L)/T(L) \sim e^{-2(b-b_1)L},\qquad b-b_1>0,
  \label{TLratio}
\end{equation}
is exponentially small in the large-$L$ limit. Thus, there is an
interval of values of $t_s$, $T_1(L) \lesssim t_s \ll T(L)$, for which
no transition to $|+\rangle$ is observed for $h\approx 0$, while a
jump to the lowest-energy positively magnetized kink-antikink state is
observed for $h\approx h_1(L)$. If instead $t_s$ is also much smaller
than $T_1(L)$, no jump is observed for $h\approx h_1(L)$: the system
is stuck in the {\em wrongly} magnetized state $|-\rangle$ also for $h
\gtrsim h_1(L)$.

\begin{figure}[!t]
 \includegraphics[width=0.9\columnwidth, clip]{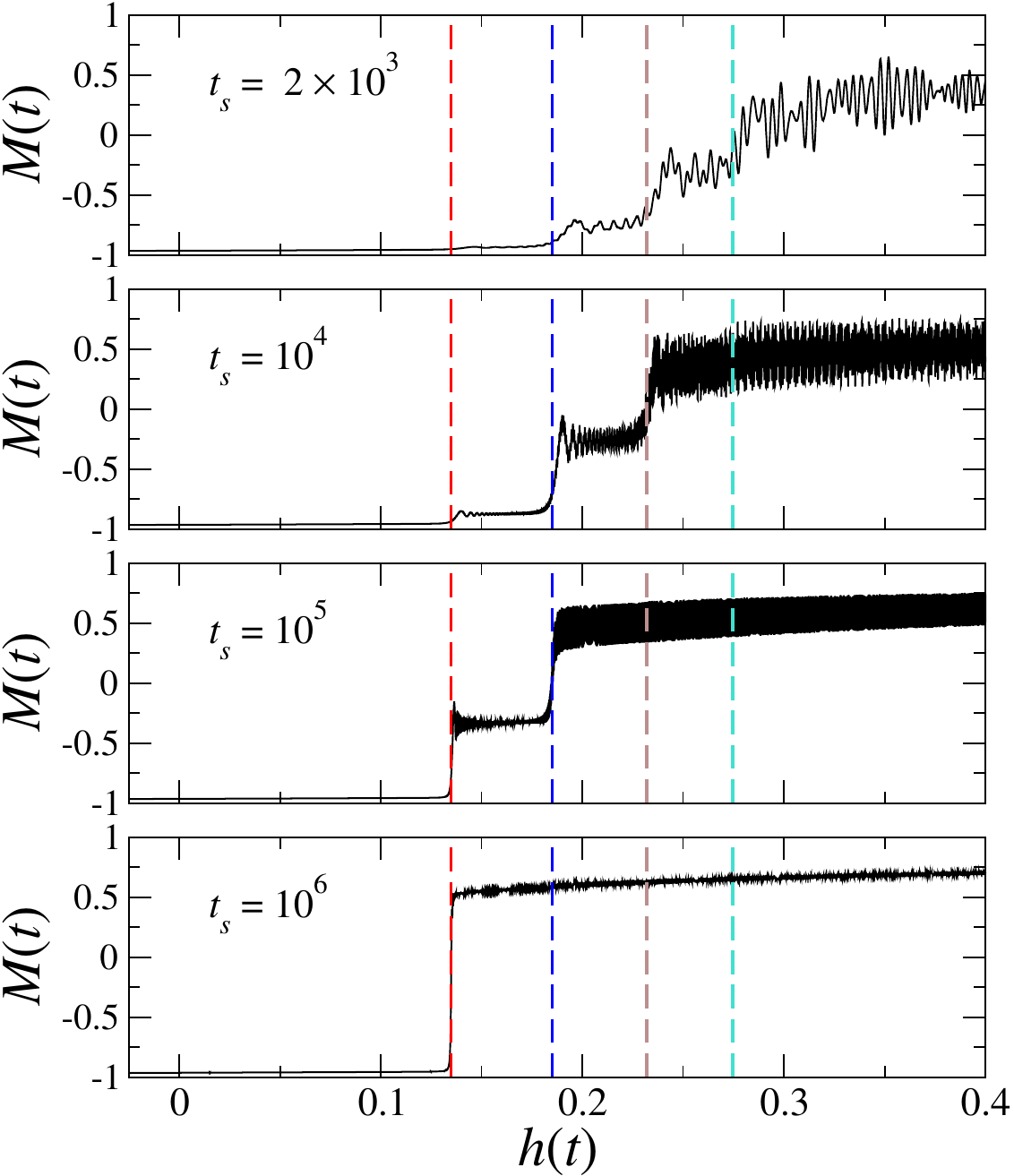}
 \caption{The magnetization in the quantum Ising chain with $L=12$
   sites as a function of $h(t)$, for the KZ protocol defined in
   Eq.~\eqref{hst}.  Each panel shows data for a different value of
   $t_s$, varying from $2 \times 10^3$ to $10^6$.  Vertical dashed
   lines denote the positions of the avoided crossings between the
   $|-\rangle$ state and the first four zero-momentum kink-antikink
   states. The corresponding time scales $T_k(L=12)$ are $T_k \approx
   3.9 \times 10^5$, $3.6 \times 10^4$, $1.0 \times 10^4$, and $4.7
   \times 10^3$, for $k=1,2,3,4$, respectively [see
     Eq.~\eqref{upsdef1}].  The time scale of the dynamics for
   $h\approx 0$ is $T(L=12) \approx 4.4 \times 10^9$, which is much
   larger than the largest $t_s$ considered.}
 \label{kzdyn1}
\end{figure}

To illustrate the previous scenario, in Fig.~\ref{kzdyn1} we show some
results for the magnetization, as obtained for $L = 12$ and some
values of $t_s\ll T(L) \approx 4.4 \times 10^9$.  We observe that, as
expected, the magnetization does not change abruptly to a positive
value for $h(t)\approx 0$, where it shows a smooth time dependence.
The behavior of the data for $h \approx h_1(L)$ (red dashed line in
Fig.~\ref{kzdyn1}), depends on $\Upsilon_1 = t_s/T_1(L)$, where
$T_1(L) = 3.9\times 10^5$.  For $t_s = 10^6 > T_1(L)$ (bottom panel),
the magnetization makes a sudden jump, indicating that the system has
essentially moved to the kink-antikink state that makes the avoided
level crossing with the negatively magnetized state. Note that
magnetization after the jump is approximately 0.5, which should be
identified with the magnetization of the kink-antikink state for such
a small value of $L$.  Equation~(\ref{magnkink}) predicts $M = m_0 (1 - 4
a_1/(3 a L^{2/3})$ for $ h = h_1(L)$. Using $m_0 \approx 0.96$, $a =
1.035$, $a_1 = 2.52$, we obtain $M = 0.37$, which is close to the
value we observe.

On the other hand, when $t_s\ll T_1(L)$ (first two top panels) no jump
is observed for $h \approx h_1(L)$. As discussed below, the
transition to a kink-antikink state occurs at one of the following
avoided level crossings.  Finally, for $t_s = 10^5 \sim T_1(L)$,
corresponding to $\Upsilon_1 = 0.256$, we observe a jump to an
intermediate value of $M$, in agreement with the two-level predictions.
If one uses $m_k \approx 0.5$, Eq.~(\ref{Masymptotico_KZ}) predicts $M
\approx -0.38$ as soon as $h(t)$ is larger than $h_1(L)$.

\section{Multiple avoided level crossings and KZ dynamics}
\label{transfinite}

The arguments and results reported in the previous section can be
extended to the avoided level crossings occurring for $h = h_k(L)$,
$k\ge 2$.  Each crossing is characterized by a different, yet
exponentially large, time scale $T_k(L)\sim L / \Delta_{m,k}(L)^{2}$,
which decreases with increasing $k$.  For a given lattice size $L$ and
time scale $t_s \ll T(L)$, there is a relevant avoided level crossing $k$
characterized by $t_s \ll T_{k-1}(L)$ and $t_s\gtrsim T_k(L)$. Under
these conditions, the system is stuck in the negatively magnetized
state $|-\rangle$ for $t < t_s h_k(L)$. At $h(t) = h_k(L)$ the system
jumps to a positively magnetized kink-antikink state.

\begin{figure}[!t]
 \includegraphics[width=0.9\columnwidth, clip]{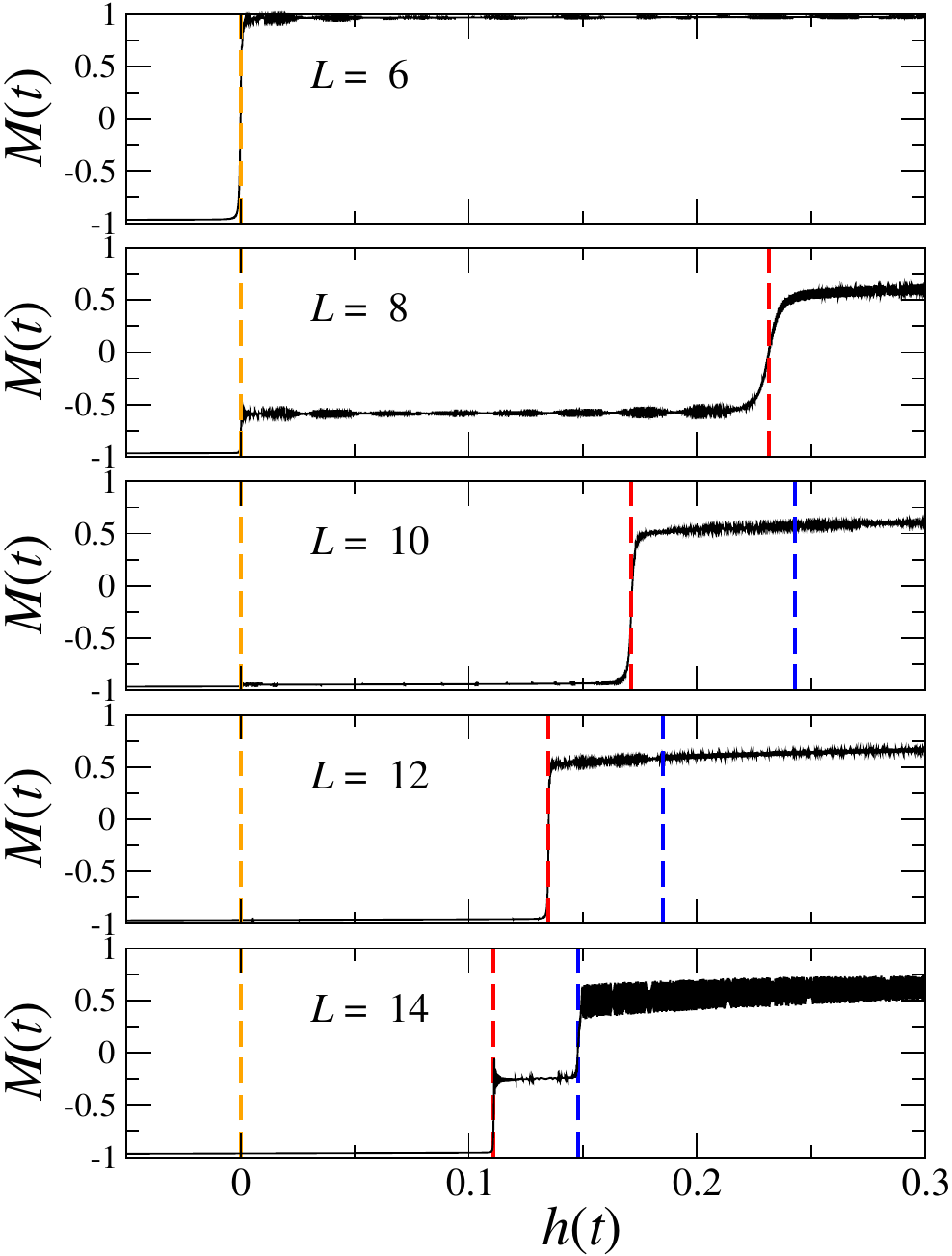}
 \caption{Time-dependence of the magnetization for the KZ protocol, at
   fixed $t_s=10^6$ and for different chain sizes ($L=6, \, 8, \, 10,
   \, 12, \, 14$, from top to bottom).  The vertical dashed lines
   indicate the position of the FOQT point $h=0$ (orange) and of the
   first avoided crossings $h_k(L)$, for $k=1$ (red) and $k=2$ (blue).
   Note that, for the smallest size $L=6$, the dynamics is adiabatic
   at the FOQT point ($h=0$), where the system exhibits a sudden jump
   from $|-\rangle$ to $|+\rangle$.  In contrast, for $L \gtrsim 8$,
   the system at $h\approx 0$ remains in the $|-\rangle$ state, until
   one of the next avoided crossings $h_k(L)$ is reached.}
 \label{kzdyn2}
\end{figure}

The above scenario is demonstrated by the numerical results of
Fig.~\ref{kzdyn2}, corresponding to $t_s = 10^6$ and different sizes
$L$. For $L=6$ (top panel), the time scale $T(L)$ is one order of
magnitude smaller than $t_s$ [$T(L=6) \approx 2.4 \times 10^5$].  As a
consequence, the system jumps to the $|+\rangle$ state for $h(t)
\approx 0$.  For $L=8$, $T(L) = 6\times 10^6$, which is not much
larger than $t_s$. Thus, the magnetization of the system is able to
make a small jump: For $h\gtrsim 0$ the system goes to a
superposition of the $|+\rangle$ state, with a small amplitude, and of
the $|-\rangle$ state. At $h(t) \approx h_1(L)$, since $t_s \gg
T_1(L)$ [$T_1(L=8) \approx 5.3 \times 10^3$], the $|-\rangle$ state is
replaced by the lowest-energy kink-antikink state.  For $L=10$ and
$12$, no jump occurs at $h = 0$, since $t_s \ll T(L)$. On the
contrary, since $t_s \gtrsim T_1(L)$ in both cases, we observe a
single jump for $h(t) = h_1(L)$.  The behavior for $L=14$ (bottom) is
more complex.  Since $t_s$ is only slightly smaller than $T_1(L) =
3.3\times 10^6$, we observe a partial jump in the magnetization,
indicating that the state is a superposition of the $|-\rangle$ state
and of the lowest-energy kink-antikink state for $h\gtrsim h_1(L)$.
Since $T_2(L=14) \approx 2.3 \times 10^5 < t_s$, the $|-\rangle$ state
disappears at the second avoided level crossing, thus, for $h(t) >
h_2(L)$, the system is mainly in a superposition of the two lowest-energy
kink-antikink states.  We observe fast oscillations (with period $T
\approx 7$, not visible in the figure) of the magnetization
which can be directly related to the
energy difference between the two levels [$2 \pi/(E_2-E_1) \approx
7.83$], indicating that the matrix element of the magnetization
between different kink states is not small.  Analogous oscillating
behaviors are observed in Fig.~\ref{kzdyn1} for $t_s \le 10^5$.

It is interesting to observe that, as $t_s$ decreases, the behavior
may become more complex, as for $t_s = 2\times 10^3$ in
Fig.~\ref{kzdyn1}.  Indeed, the system may perform several small
jumps, becoming eventually a superposition of several kink-antikink
states. Moreover, the level spacing of the kink-antikink states
decreases as $k$ increases (for $g$ small $E_{k+1}(L) - E_{k}(L)
\approx c_k L^{-2/3}$ with $c_k\sim k^{-1/3}$, see App.~\ref{AppA.2})
while the gap $\Delta_{k,m}(L)$ increases with $k$. Thus, we may end
up in a situation in which the two-level truncation is no longer valid,
with several states involved in the crossing.  Moreover, as $L$
increases, the kink-antikink levels get closer, so this phenomenon may
become more pronounced for large $L$.  In the same limit, the
different avoided crossings also become closer and closer, since
$h_{k+1}(L)-h_k(L)\sim L^{-5/3}$, eventually collapsing towards the
first crossing $h_1(L)\approx a/L$, which might be interpreted as a
size-dependent spinodal point.

We mention that a similar qualitative behavior for the KZ dynamics in
finite-size Ising chains was also highlighted in Ref.~\cite{SCD-21},
where it was interpreted as a phenomenon of quantized
nucleation. Note, however, that our analysis provides a more thorough
description of these out-of-equilibrium phenomena at the FOQTs of
quantum Ising models. Indeed, we provide an exact numerical
characterization, which is confirmed by an exact perturbative
computation, of the size dependence of the values $h_k(L)$ of the
longitudinal magnetic field where the crossings occur, see
Eq.~\eqref{hkbeh}. In particular, this implies their collapse onto a
single (spinodal) point in the limit in which the magnetic energy $hL$
is kept fixed as $L\to\infty$. Finally, we provide general EFSS and
OFSS theories for the avoided crossings, which allows us to provide a
quantitative estimate of the time scales $T_k(L)$.  These results are
crucial to quantitatively interpret how the observed quantum evolution
depends on $t_s$.

\section{The KZ protocol in the thermodynamic limit}
\label{transLinfty}

We finally turn to the investigation of the out-of-equilibrium
behavior arising in the KZ evolution in the thermodynamic limit.  The
analysis presented in the previous sections is only appropriate for
finite-size systems for which $hL$ is fixed and not too large.  On the
other hand, if we consider the behavior for $L \to \infty$ and fixed
magnetic field, the avoided level crossings collapse towards $h=0$.
Therefore, if a scaling behavior occurs in this limit, it may not be
described by a straightforward extension of the OFSS results.

In the absence of a theoretical framework, we investigate the KZ
dynamics in the infinite-size limit numerically.  Unfortunately, even
after exploiting momentum conservation, the numerics based on exact
diagonalization is limited to relatively small system sizes, up to a
few tens of sites.  Nevertheless, as we shall see, an appropriate
analysis of the finite-size data allows us to unveil a scaling picture
of the behavior of the KZ dynamics in the thermodynamic limit.

To study the KZ protocol in the infinite-size limit, we follow the
following two-step procedure: (i) we first determine the large-$L$
limit at fixed time scale $t_s$, by increasing $L$ until the relevant
observables $O(t,t_s,L)$ appear to approach $L$-independent quantities
$O_\infty(t,t_s)$, as a function of $h(t)=t/t_s$; (ii) we study the
behavior of the infinite-size limits $O_\infty(t,t_s)$ as a function
of $t_s$, looking for a scaling behavior in terms of the variables $t$
and $t_s$, for large values of $t_s$.

\begin{figure}[!t]
  \includegraphics[width=0.9\columnwidth, clip]{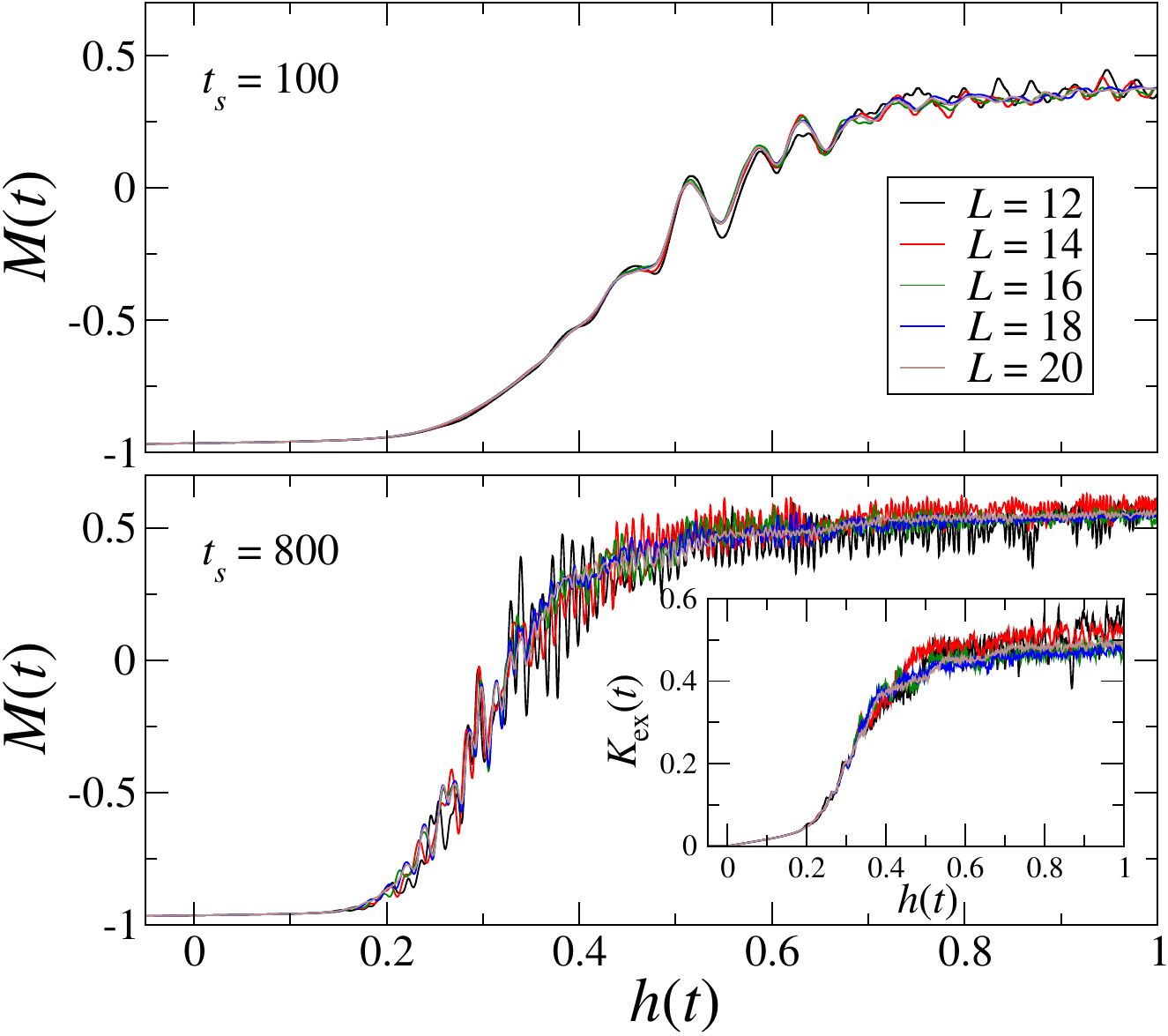}
  \caption{Evolution of the magnetization $M(t)$ in the KZ protocol,
    for $t_s=100$ (top) and $t_s=800$ (bottom), and for various values
    of $L$ (see legend). Data are plotted versus $h(t) = t/t_s$. The
    inset in the bottom panel shows the time behavior of the excess
    energy $K_{\rm ex}(t)$ defined in Eq.~(\ref{eneexdef2}).  In all
    cases the data appear to approach a well defined large-$L$ limit.}
    \label{thermodlimts100}
\end{figure}

The procedure corresponding to step (i) is exemplified in
Fig.~\ref{thermodlimts100}, which displays the time evolution of the
longitudinal magnetization $M(t)$ defined in Eq.~\eqref{mxm} for
$t_s=100$ (top) and $t_s=800$ (bottom), and various values of $L$.
For $t_s=800$, we also report data for the excess energy $K_{\rm
  ex}(t)$ defined in Eq.~\eqref{eneexdef2}.  We observe that the
different data sets apparently converge to an asymptotic large-$L$
behavior, which provides the time dependence of the infinite-size
magnetization $M_\infty(t,t_s)$ and of the excess energy $K_{{\rm ex},
  \infty} (t,t_s)$ at the given value of $t_s$.  Convergence is faster
for small time scales, as is visible from the top panel.  In contrast,
when $t_s$ increases, fast oscillations in time emerge, although their
amplitude decreases with $L$ and a qualitative asymptotic behavior can
be recognized for moderate chain lengths $L \approx 20$ up to
$t_s=O(10^3)$ (bottom panel).  The behavior of the excess energy
$E_{\rm ex}(t)$ defined in Eq.~\eqref{eneexdef} (not shown) resembles
that of $K_{\rm ex}(t)$, although with wider time oscillations, which
are induced by the transverse-field energy contribution.

\begin{figure}[!t]
 \includegraphics[width=0.9\columnwidth, clip]{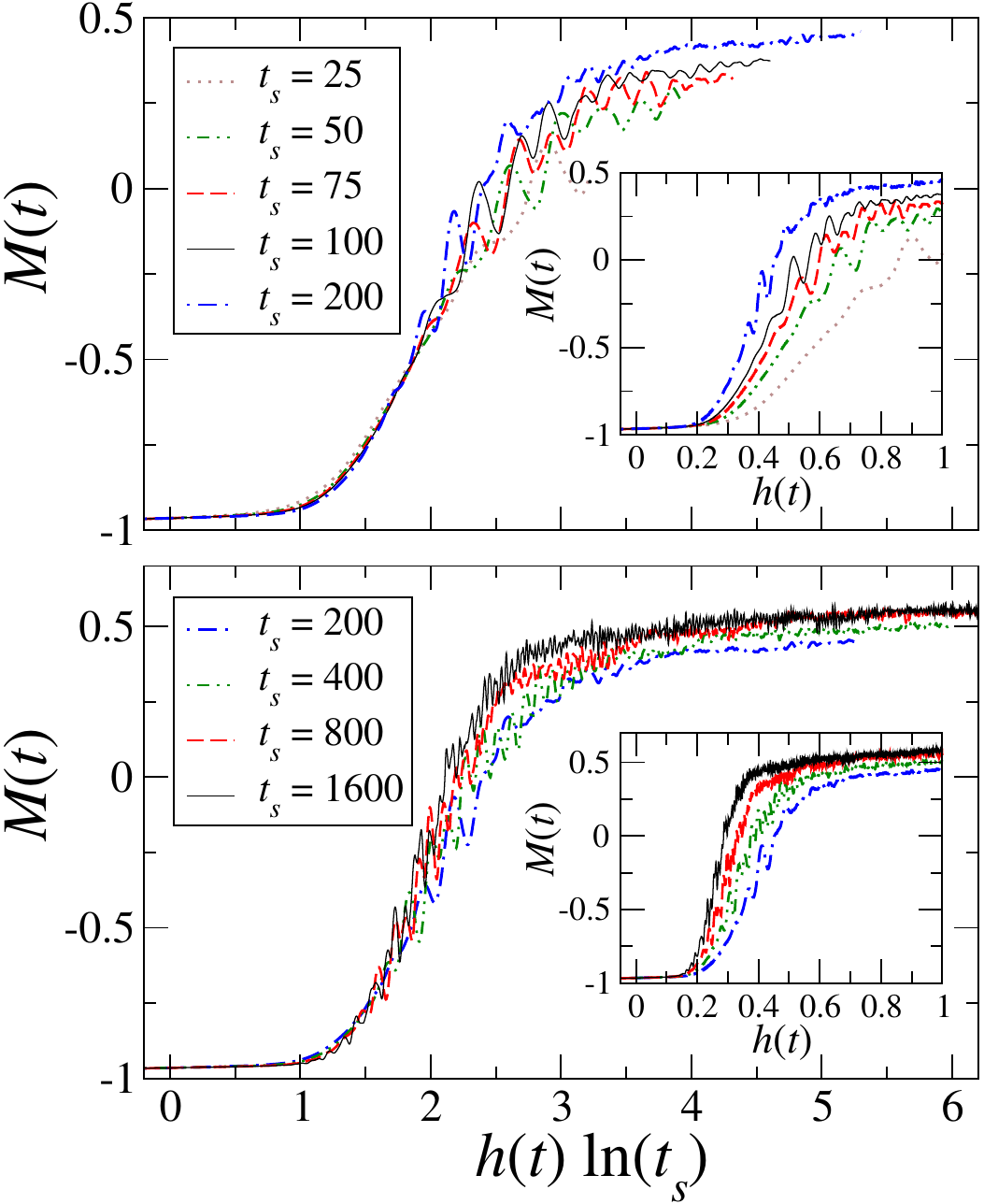}
 \caption{The infinite-size magnetization for different KZ
   protocols, plotted versus $\Omega \equiv h(t)\ln t_s$ (main frames)
   and versus $h(t)=t/t_s$ (insets).
   Data are shown for smaller values of $t_s$, from $25$ to $200$ (top)
   and for larger values of $t_s$, from $200$ to $1600$ (bottom).
   The dynamics starts from $h_i = -0.05$ in all cases and is for
   systems of size $L=20$, ensuring that the large-$L$ limit
   is approached for all the considered values of $t_s$.
   Note that the curves tend to collapse on a single curve, when they
   are plotted versus $\Omega$. This suggests the existence of a
   nontrivial large-$t_s$ scaling limit in terms of $\Omega$.}
 \label{thermodlimmag}
\end{figure}

\begin{figure}[!t]
 \includegraphics[width=0.9\columnwidth, clip]{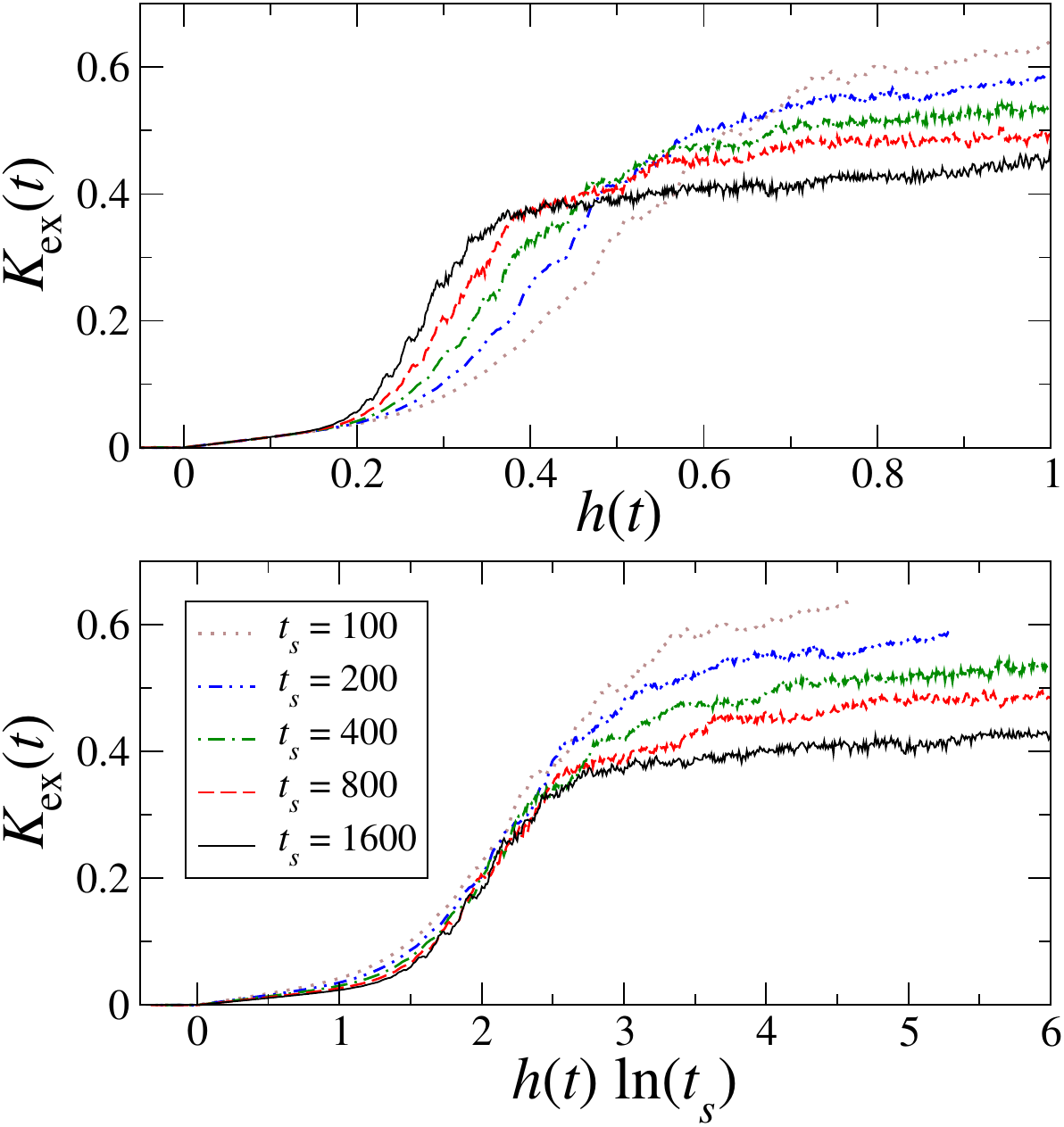}
 \caption{The excess energy $K_{\rm ex}$, as defined in
   Eq.~(\ref{eneexdef2}), for the same KZ evolutions reported in
   Fig.~\ref{thermodlimmag}.  Data are for $L=20$, ensuring that the
   large-$L$ limit is obtained.  As for the magnetization, we observe
   an apparent data collapse, when they are plotted versus $\Omega$.
   The data show a slower approach to the scaling
   curve in the large-$\Omega $ region, $\Omega \gtrsim 4$, where the
   magnetization is almost constant and independent of $t_s$ ($M\approx
   0.5$, for these values of $\Omega$, see Fig.~\ref{thermodlimmag}).}
 \label{thermodlimexc}
\end{figure}

In step (ii) we compare the infinite-size time evolutions of the
observables for several increasing time scales $t_s$, looking for the
emergence of a scaling behavior. The two insets of Fig.~\ref{thermodlimmag}
and the top panel of Fig.~\ref{thermodlimexc} show some results
for the longitudinal magnetization $M(t)$ and for the excess energy
$K_{\rm ex}(t)$ versus $h(t)$, respectively.  The different curves
correspond to values of $t_s$ from $25$ to $1600$, spanning two orders
of magnitude. For these values of $t_s$, we have a sufficiently
good numerical evidence that the time
evolutions obtained for $L=20$ (see Fig.~\ref{thermodlimts100}) can be
effectively considered as infinite-size behaviors.  We note that the
magnetization changes sign for values of $h$ that decrease with
increasing $t_s$, suggesting that the sign change occurs for values of
$h$ that approach $h(t)=0$ (thus $t=0$) as $t_s\to\infty$.

The plots reported in the two main frames of Fig.~\ref{thermodlimmag}
and in the lower panel of Fig.~\ref{thermodlimexc} show an apparent collapse of 
the infinite-size magnetization and excess-energy curves when plotted versus
\begin{equation}
  \Omega(t) = h(t) \, \ln t_s = {t \ln t_s\over t_s}.
  \label{Xidef}
\end{equation}
Therefore, they suggest a peculiar infinite-size scaling behavior for
large values of $t_s$, that is
\begin{eqnarray}
M_\infty(t,t_s) &\approx& m_0 \,{\cal M}_\infty(\Omega), \label{miscai}\\
K_{{\rm
      ex},\infty}(t,t_s) &\approx& {\cal K}_{{\rm ex},\infty}(\Omega) .
\label{kiscai}
\end{eqnarray}
Notice that $\Omega$ definitely differs from the naive large-size
scaling variable $W$ that is derived in the OFSS framework,
cf. Eq.~(\ref{Wdef}).  Equation~(\ref{miscai}) predicts that the
longitudinal magnetization changes sign for a fixed value of $\Omega$
and therefore for $h(t)\sim 1/\ln t_s$, somehow resembling a
spinodal-like behavior.  Further studies are required to make this
scenario more solid.

We conclude by noting that the magnetization, displayed in
Fig.~\ref{thermodlimmag}, is almost constant (apart from short-time
fluctuations), $M_\infty(t\to\infty,t_s)\approx 0.5$, for sufficiently
large values of $\Omega$ and of $t_s$.
The significant deviation from 1, which is
the ground-state value for $h\to\infty$, can be explained by the large
energy excess, therefore by the fact that the KZ protocol has injected
a relatively large amount of energy (work to change $h$) in the
system, giving rise to a significant departure from the ground state
of the Hamiltonian at large times.  This is clearly evident in the
behavior of the excess energy $E_{\rm ex}$ (not shown), which is
qualitatively analogous to the curves of $K_{\rm ex}$ reported in
Fig.~\ref{thermodlimexc}.

\begin{figure}[!t]
 \includegraphics[width=0.9\columnwidth, clip]{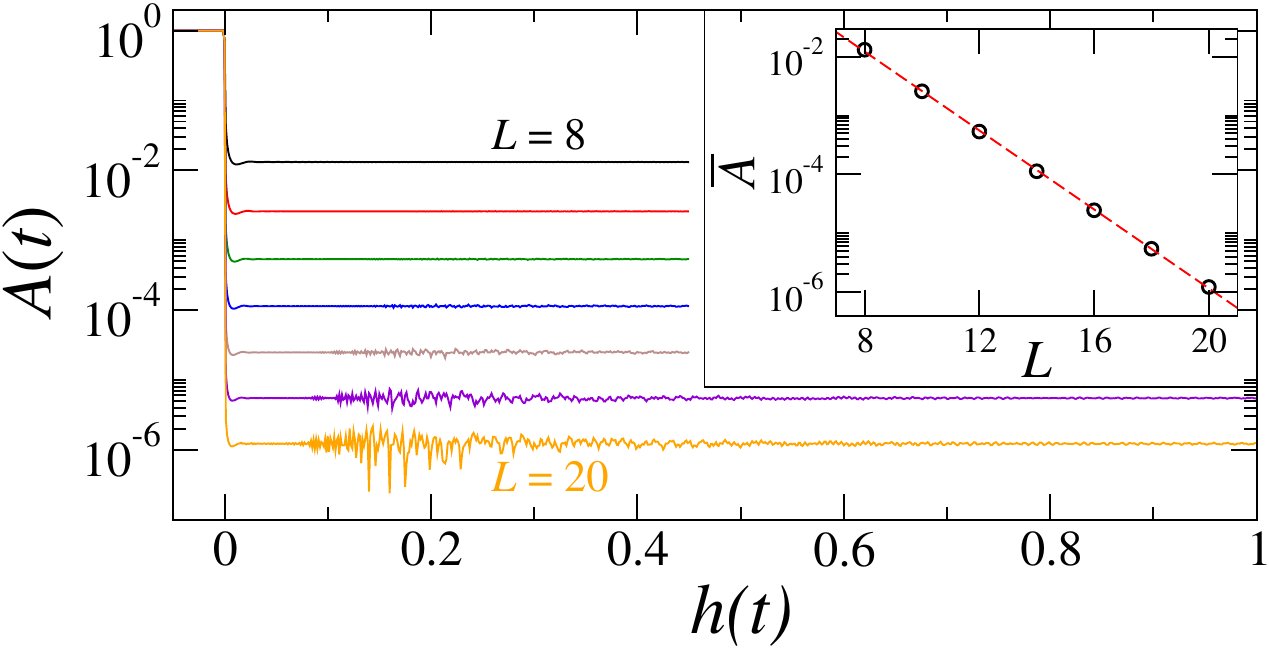}
 \caption{The adiabaticity function $A(t)$ for KZ protocols starting
   from $h_i = -0.05$ and with a time scale $t_s = 800$.  Results for
   different system sizes $L=8, \, 10, \, 12, \, 14, \, 16, \, 18, \,
   20$ (top to bottom curves).  The inset shows the value $\bar A$
   obtained by averaging the numerical data for $h(t) \in [0.5,1]$;
   the straight line is an exponential fit $\bar{A} = A_0 \exp(-\alpha
   L)$, with $\alpha \approx 0.772$.}
 \label{fig:overlap}
\end{figure}

To gain further insight on the features of such stationary states, in
Fig.~\ref{fig:overlap} we focus on the adiabaticity function defined
in Eq.~\eqref{adiabfunc}.  For $h \approx 0$, we observe a sudden drop
of $A(t)$ to a stationary value $\bar A$ close to zero, meaning that,
as soon as the FOQT point is crossed, the state of the system becomes
nearly orthogonal to the instantaneous ground state at the
corresponding field $h(t)$. As discussed in the previous sections,
this happens as soon as the KZ time scale $t_s$ is much smaller than
$T(L)$, the characteristic time associated with the avoided level
crossing at $h=0$. Indeed, $T(L)$ varies from $T(L\!=\!8) \approx 7.28
\times 10^6$ to $T(L\!=\!20) \approx 8.36 \times 10^{14}$, and is in
all cases much larger than $t_s = 800$, the value used in the KZ
dynamics considered in the figure.  We remark that the asymptotic
stationary value $\bar A$ is almost insensitive to the choice of the
time scale $t_s \ll T(L)$, while it decreases exponentially with the
system size $L$, as reported in the inset.  In particular, a numerical
fit of our data obtained with $t_s=800$ gives $\bar A(L) \sim
e^{-\alpha L}$, with $\alpha \approx 0.772$.

\begin{figure}[!t]
 \includegraphics[width=0.9\columnwidth, clip]{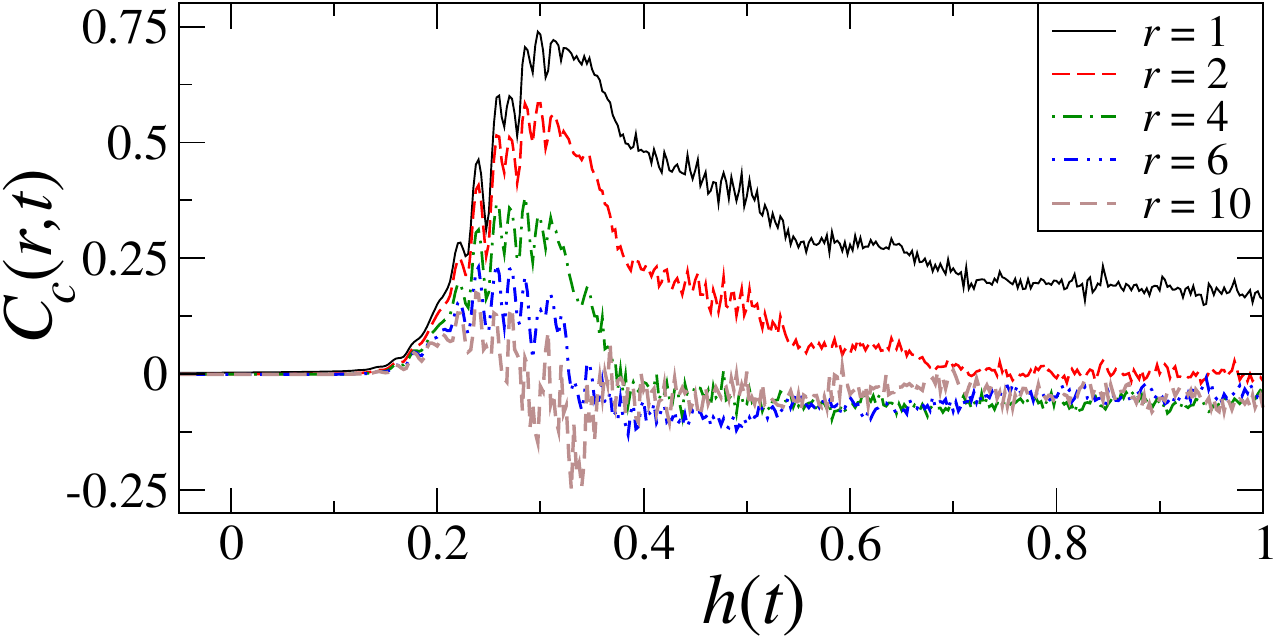}
 \caption{Connected correlation function of the longitudinal
   magnetization $C_c(r,t)$, evaluated at various distances $r \leq
   L/2$, for a chain of length $L=20$. Results are for the same KZ
   evolutions considered in Fig.~\ref{fig:overlap}. }
 \label{fig:correlations}
\end{figure}

Finally, in Fig.~\ref{fig:correlations} we show the connected part of
the correlation function of the longitudinal magnetization, for a
system with $L=20$ spins. Assuming translational invariance, it is
defined as
\begin{equation}
  C_c(r,t) = \langle \Psi(t) | \sigma^{(1)}_x \sigma^{(1)}_{x+r} | \Psi(t)\rangle
  - \langle \Psi(t) | \sigma^{(1)}_x | \Psi(t)\rangle^2.
\end{equation}
Our data are plagued by oscillations in time, whose frequency
increases with $t_s$. Such oscillations decrease for larger system
sizes, which, unfortunately, we were only able to increase up to $L
\approx 20$.  Nonetheless we observe that, apart from the
intermediate-time transient region around $h(t) \approx 0.3$,
corresponding to the time at which the magnetization changes sign
(compare with the dashed red curve for $t_s=800$, in the inset of
the bottom panel of Fig.~\ref{thermodlimmag}),
the system is always weakly spatially
correlated and the correlation length is of the order of one.

\begin{figure}[!t]
 \includegraphics[width=0.9\columnwidth, clip]{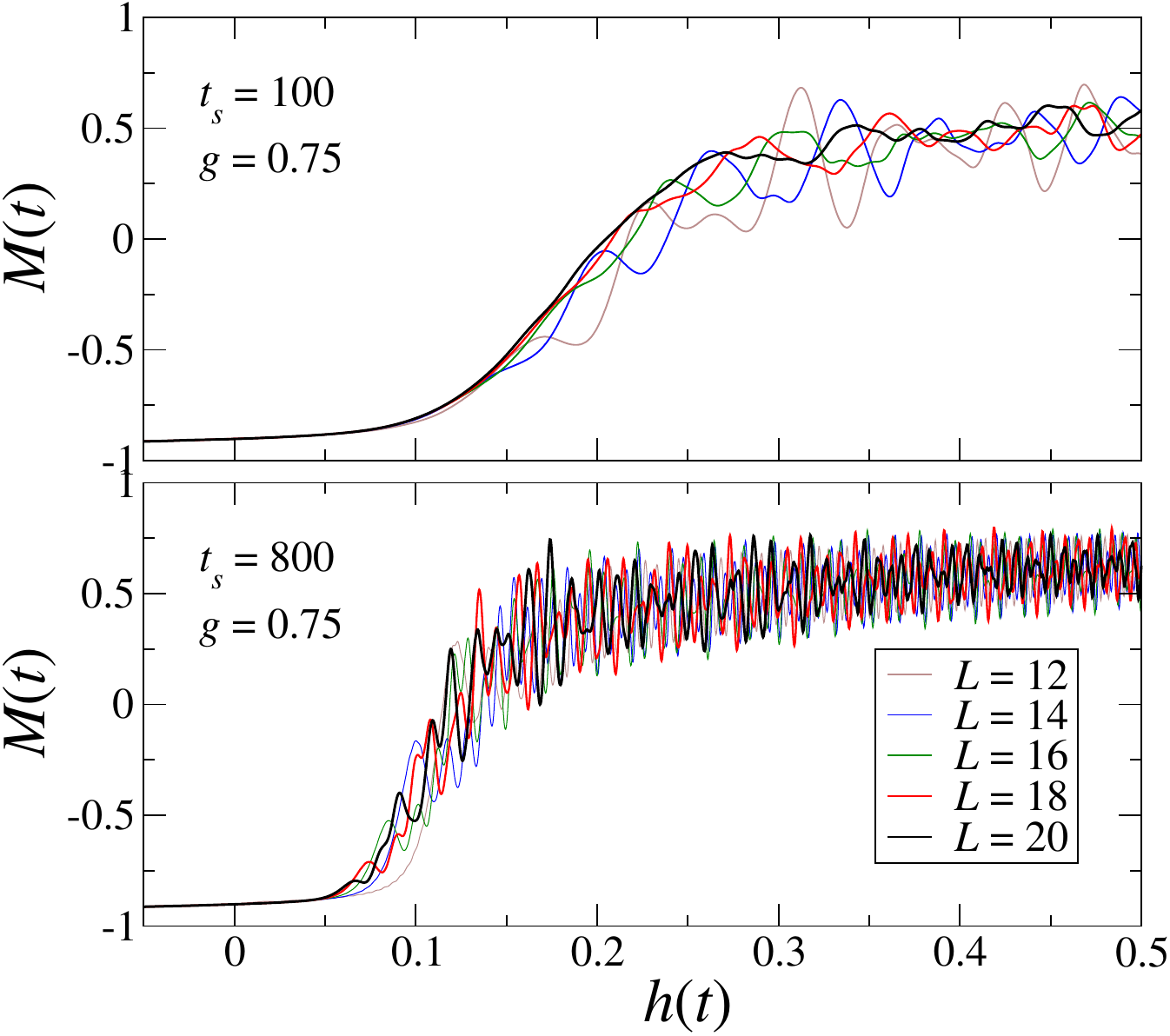}
 \caption{Same plots of the infinite-size magnetization as in
   Fig.~\ref{thermodlimts100}, but for $g=0.75$.
   Note the appearance of more marked oscillations, compared to
   the $g=0.5$ case, especially when increasing $t_s$.
   This suggests that larger system sizes than those considered here
   are required to observe the large-$L$ limit.}
 \label{thermodlimts100_2}
\end{figure}

\begin{figure}[!t]
 \includegraphics[width=0.9\columnwidth, clip]{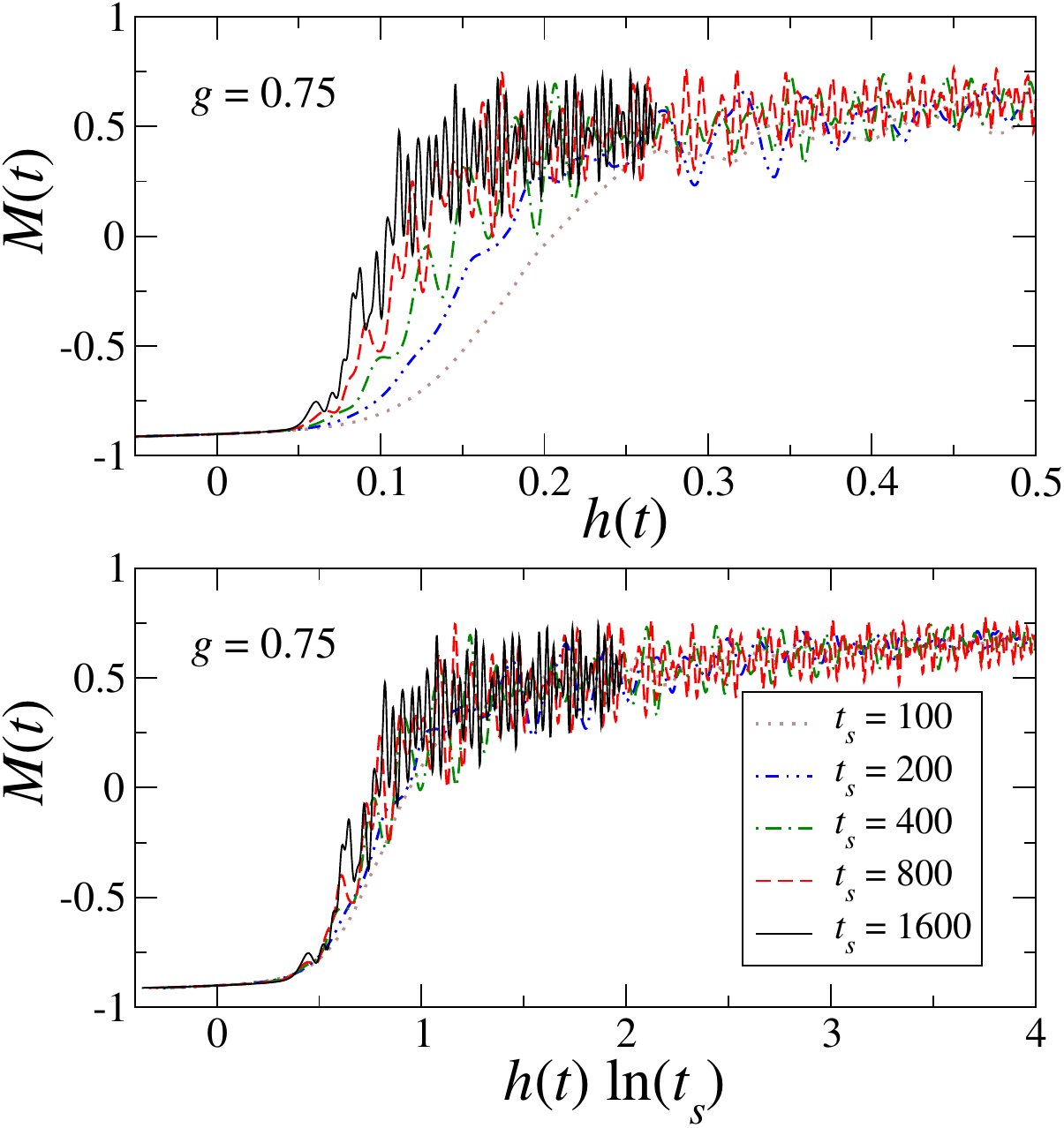}
 \caption{Same plots as in Fig.~\ref{thermodlimmag}, but for $g=0.75$
   and for $t_s \geq 100$.
   We show data up to smaller values of $h(t)$, as here the magnetization
   becomes positive for smaller field strengths.}
 \label{thermodlimmag2}
\end{figure}

We should stress that the numerical analysis reported above
has been performed after fixing the transverse-field strength at $g=0.5$.
Nonetheless, analogous behaviors, characterized by the same scaling function
$\Omega$ defined in Eq.~(\ref{Xidef}), are expected to occur along
the whole FOQT line of the Ising chain (i.e., for any $g<1$).
To support universality of the infinite-size KZ behavior,
we have performed extensive simulations also for other values of $g$,
both above and below $g=0.5$.
The results shown in Figs.~\ref{thermodlimts100_2} and~\ref{thermodlimmag2},
which refer to $g=0.75$, support this conjecture.
Unfortunately, evidence of such KZ scaling would require larger
time scales $t_s$ and larger
and larger chains, when approaching the continuous transition point $g=1$,
since the infinite-volume correlation length $\xi_m$ in the magnetized
phases diverges as $(1-g)^{-1}$ in this limit.

This scenario clearly emerges from the results of Fig.~\ref{thermodlimts100_2},
showing the magnetization $M$ as a function of $h(t)$, for different
values of $L$. In fact, when $t_s$ increases, the appearence of pronounced
time oscillations (more evident than those for $g=0.5$---compare with
Fig.~\ref{thermodlimts100}) prevents us from identifing a clear
asymptotic large-$L$ behavior, although we find robust evidence of a
converged global growth of the various curves.
Such oscillations should be ascribed to revival phenomena, due to the
finite size of the system, and are expected to increase when
approaching the quantum critical point (see, e.g.,
Refs.~\cite{HHH-12, KLM-14, Cardy-14, JJ-17, MAC-20, RV-20}).
A fairly good data collapse of the curves at the largest available
size $L=20$ and for sufficiently large values of $t_s \geq 100$,
when plotted versus $\Omega(t)$, is visible in Fig.~\ref{thermodlimmag2}.
We mention that we recover a similar behavior also for $g=0.4$, where
time oscillations are less evident (not explicitly shown here).

\section{Summary and conclusions}
\label{conclu}

We have addressed the out-of-equilibrium dynamics of quantum Ising
models in a transverse field, driven across their FOQTs by a
homogeneous longitudinal external field $h$. Specifically, we focus on
the quantum spin chain with Hamiltonian (\ref{hedef}) and periodic
boundary conditions, which provides a paradigmatic model for which we
can obtain accurate numerical results that can be used to verify our
theoretical predictions.  We consider a dynamic KZ protocol in which
the longitudinal field $h$ is varied linearly in time, with a time
scale $t_s$, i.e, $h(t)=t/t_s$, across the FOQT at $h=0$.
The system starts from the negatively magnetized ground state
at $h_i\equiv h(t_i)<0$, and then it evolves unitarily according to
the Schr\"odinger equation for $t>t_i$, up to positive values of
$h(t)$, eventually leading to states with positive longitudinal
magnetization.  We analyze the time evolution of some relevant
observables, such as the longitudinal magnetization and the excess
energy, identifying scaling regimes involving the system size $L$ and
the time scale $t_s$ of the KZ protocol.  We also discuss the behavior
in the infinite-size limit, keeping $t_s$ fixed.

In finite-size systems, we identify an OFSS regime whenever the
system is close to one of the avoided level crossings that
characterize the spectrum in the presence of the longitudinal field
$h$.  In particular, at the FOQTs for $g<1$, since periodic boundary
conditions preserve the ${\mathbb Z}_2$ symmetry of the model, an
avoided level crossing occurs for $h=0$, where the actual eigenstates
are superposition of the positively and negatively magnetized states,
with an energy gap $\Delta(L)$ that decays exponentially with $L$,
i.e., $\Delta(L)\sim e^{-bL}$ with $b \approx -\ln g$.  The quantum
evolution of finite-size systems driven across $h=0$ shows an OFSS
behavior~\cite{PV-24,RV-21,PRV-18-def,PRV-20}, when $L\to\infty$ and
$t_s\to\infty$ simultaneously, keeping appropriate combinations of
$L,t_s,t$ fixed.  A crucial scaling variable is the time-scale ratio
$\Upsilon = t_s/T(L)$, where $T(L) \sim L \Delta(L)^{-2}$ is the time
scale associated with the passage across the avoided level crossing.
In particular, if $t_s\gg T(L)$ the system evolves adiabatically, thus
switching from the negatively magnetized state to the positively
magnetized one as it crosses $h=0$.  On the other hand, if $t_s\ll
T(L)$, the passage across $h=0$ is effectively instantaneous, so that
the system persists in the {\em wrongly} magnetized state ($M<0$) for
$h(t)>0$.

Besides the avoided crossing at $h=0$, the spectrum of
finite-size systems shows a sequence of avoided level crossings
between the ${\rm wrongly}$ magnetized state and a discrete series of
zero-momentum kink-antikink states, labeled by $k=1,2, \ldots$.  These
avoided crossings are localized at $h_k(L) = a/L + a_{1k}/L^{5/3} +
O(L^{-2})$, where the leading term does not depend on $k$, so that
they get closer and closer with increasing $L$, as $h_{k}-h_{k'} \sim
L^{-5/3}$.  The minimum energy differences $\Delta_{m,k}$ at each
avoided crossing are exponentially small with increasing $L$, i.e.,
$\Delta_{m,k}\sim e^{-b_k L}$. Moreover, $\Delta_{m,k}(L)$ increases
with $k$, $\Delta_{m,k}(L) < \Delta_{m,k+1}(L)$, and is always larger
than $\Delta(L)$.  These predictions have been verified numerically
for $k$ up to 5 on systems of size up to $L=26$.  We conjecture that
these features also hold for higher excited states, for sufficiently
low energies, smaller than the values where qualitatively different
states appear (for example, we must require at least $k/L\lesssim
1/2$).

The dynamic behavior across these additional avoided crossings
is analogous to that discussed for the avoided level crossing at
$h=0$. They also admit an OFSS description in terms of the analogous
scaling variable $\Upsilon_k = t_s/T_k(L)$, where $T_k(L)\sim L
\Delta_{m,k}(L)^{-2}$.  Such OFSS behavior can be observed because
it occurs in a very small range of values of $h$ around $h_k(L)$,
i.e., for $|h-h_k(L)|\sim \Delta_{m,k}(L)/L$, much smaller than
the $O(L^{-5/3})$ spacing between subsequent level crossings.
Note that $T(L) \gg T_1(L) \gg T_2(L) \gg \ldots$, since
$\Delta(L) \ll \Delta_{m,1}(L) \ll \Delta_{m,2}(L) \ldots$.

The previous results allow us to predict the behavior of the system
along the KZ evolution.  According to the OFSS theory, if the system
starts in the ground state for $h < 0$, i.e., in the negatively
magnetized state, it may jump to a state with positive magnetization
at $h_k(L)$ only if $t_s \gtrsim T_k(L)$.  Since the time scales
satisfy $T(L) \gg T_1(L) \gg T_2(L) \gg \ldots$, the
time scale $t_s$ of the KZ protocol can be tuned in such a way to
select at which avoided crossing the system magnetization changes
sign.  More precisely, when $t_s$ is large but still satisfies $t_s\ll
T(L)$, the negatively magnetized state effectively survives across the
$h=0$ and the first $k-1$ avoided level crossings up to the one
satisfying $t_s\approx T_k(L)$. When $h(t) \approx h_k(L)$, then the
system jumps to a kink-antikink state with positive magnetization.  We
have shown that this scenario is actually realized, by reporting
numerical results for the quantum evolution of systems along KZ
protocols up to $L = 20$ and $t_s = 10^6$.

As already stressed, the avoided level crossings related to
kink-antikink bound states collapse to $h=0^+$ in the infinite-size
limit. This makes the study of the thermodynamic limit (defined as the
limit $L\to\infty$ keeping $t$ and $t_s$, and therefore $h(t)$,
constant) problematic.  We study such limit numerically, by first
determining the large-volume behavior at fixed $t_s$, and then
analyzing the behavior with increasing $t_s$, looking for the scaling
behavior that characterizes the out-of-equilibrium behavior of the
infinite-size systems across the FOQT.
Our analysis shows that the negatively magnetized state jumps to
states with positive magnetization at values $h_\star(t_s)>0$ that
approach $h=0^+$ with increasing $t_s$.  On the basis of the numerical
results, we conjecture that $h_\star(t_s)\sim 1/\ln t_s$, which
suggests an infinite-size scaling behavior in terms of the scaling
variable $\Omega = t \ln t_s/t_s$.  Another notable feature of the
dynamics is that the quantum Ising system approaches a nontrivial
stationary state in the large-$t$ limit, characterized by an energy
significantly larger than that of the corresponding magnetized ground
state. This large energy difference is clearly related to the average
work done when varying $h$ in the KZ protocol.  It is important to
stress that this infinite-size behavior is not obtained by a
straightforward extension of the finite-size analyses based on the
OFSS behaviors across the avoided level crossings.  Of course,
additional computations are called for to conclusively support the
above infinite-volume scenario.  In particular, to improve the
evidence of the scaling behavior, results for larger sizes would be
highly desirable.

We finally remark that the spinodal-like behavior observed here shows
notable similarities with that observed when short-range classical
systems are driven across a first-order transition by varying a
Hamiltonian parameter with time, even if the dynamics is different:
results with a purely relaxational dynamics are reported, e.g., in
Refs.~\cite{PV-17,PV-25}.

Although the above results have been obtained for quantum Ising chains
with periodic boundary conditions, we believe that several features
have general validity.  For example, apart from specific details, the
general scenario for finite-size and infinite-size systems should
apply to open and parallel fixed boundary conditions. This is
desirable from a numerical point of view, since one could employ
methods, such as the density-matrix renormalization group, which are
best suited to simulate chains with open ends and which would enable
us to obtain results for larger system sizes, potentially up to
$L\gtrsim 100$.  Some differences are expected when considering
antiperiodic and opposite fixed boundary conditions, since the
lowest-energy states are kink
states~\cite{CNPV-14,PRV-18-fowb,PRV-20}, separating regions with
different magnetization. It would be tempting to understand how the
infinite-volume out-of-equilibrium behavior is realized there.

We believe that qualitatively analogous scaling behaviors, in
finite-size and infinite-size systems, should also emerge in higher
dimensions, such as in the two- and three-dimensional quantum Ising
models.  Another interesting issue is related to the role of
dissipation, which can be introduced by using, e.g., the Lindblad
framework~\cite{RV-21,DRV-20}.

We point out that the out-of-equilibrium scaling behaviors
reported in this paper have been numerically observed in relatively
small systems.  Therefore, given the need for high accuracy without
necessarily reaching scalability to large sizes, it would be tempting
to employ the available technology to check these predictions 
experimentally,
using, for instance, ultracold atoms in optical
lattices~\cite{Bloch-08, Simon-etal-11}, trapped
ions~\cite{Edwards-etal-10, Islam-etal-11, LMD-11, Kim-etal-11,
  Richerme-etal-14, Jurcevic-etal-14, Debnath-etal-16}, as well as
Rydberg atoms in arrays of optical microtraps~\cite{Labuhn-etal-16,
  Guardado-etal-18, Keesling-etal-19, BL-20} or even quantum computing
architectures based on superconducting qubits~\cite{Barends-etal-16,
  Gong-etal-16, CerveraLierta-18, Ali-etal-24}.  Some recent
experiments have already addressed the dynamics and the excitation
spectrum of quantum Ising-like chains~\cite{Gong-etal-16, LTDR-24,
  DMEFY-24}, thus opening possible avenues where the envisioned
behaviors at FOQTs can be observed in the near future.

\bigskip

\begin{center}
  {\bf DATA AVAILABILITY}
\end{center}

\medskip

The data that support the findings of this article are not publicly
available. The data are available from the authors upon reasonable
request.

\appendix

\section{Kink-antikink states:
  scaling analysis of the energies and magnetizations}
\label{kkbound}

In this Appendix we compute the energy and the magnetization of the
kink-antikink states for a chain of $L$ spins, to first order in the
transverse magnetic field $g$. We mainly consider periodic boundary
conditions, but we also address systems with fixed parallel boundary
fields.  We obtain exact results for the energies of kink-antikink
states for finite values of $L$, generalizing the results of
Refs.~\cite{MW-78,Rutkevich-08,Coldea-etal-10,Rutkevich-10}. We then
obtain scaling results for small values of $h$ and large values of
$L$, more precisely, for $g$ small and in the finite-size scaling
regime in which $h\to 0$ and $L\to \infty$ simultaneously, keeping the
magnetic energy $hL$ fixed.  Phenomenologically, the scaling regime
occurs when $h/g$ is small, but still satisfies the condition $h/g \gg
L^{-3}$.  Note that the perturbative approach also requires the
magnetic energy $hL$ to be small with respect to the spacing $4 J$ of
the levels for $g=h=0$, thus our results hold only for $h\ll 4/L$ (we
recall that $J=1$).

\subsection{Secular equation}

The spectrum of the model has been discussed in Sec.~\ref{spectrum}.
For $g = h =0$ the kink-antikink states are degenerate with energy
$E_k = 4 J + E_{gs}$, where $E_{gs}$ is the energy of the fully
magnetized lowest-energy degenerate states. To determine the energy
for $g\not=0$, we note that, because of the periodic boundary
conditions, the Hamiltonian is translation invariant. This implies
that the Hamiltonian restricted to the subspace spanned by the
kink-antikink states is the sum of $L$ blocks, each of them of size
$L-1$.  Each block is specified by a momentum $p = 2 \pi k/L$, with $k
= 0,\ldots, L-1$, with normalized basis (we write it explicitly for
$L=4$):
\begin{eqnarray}
|p,1\rangle &=& \tfrac12 \left( |-1,1,1,1\rangle + e^{ip} |1,-1,1,1\rangle + 
   \right. \label{App:basis} \\
   && \quad \left.
    e^{2ip}\ |1,1,-1,1\rangle + e^{3ip}\ |1,1,1,-1\rangle \right), \nonumber  \\
|p,2\rangle &=& \tfrac12
  \left( |-1,-1,1,1\rangle + e^{ip}\ |1,-1,-1,1\rangle + 
   \right. \nonumber \\
   && \quad \left.
    e^{2ip}\ |1,1,-1,-1\rangle+e^{3ip}\ |-1,1,1,-1\rangle \right), \nonumber  \\
|p,3\rangle &=& \tfrac12
    \left( |-1,-1,-1,1\rangle + e^{ip}\ |1,-1,-1,-1\rangle + 
   \right. \nonumber \\
   && \quad \left.
    e^{2ip}\ |-1,1,-1,-1\rangle + 
      e^{3 ip}\ |-1,-1,1,-1\rangle\right). \nonumber 
\end{eqnarray}
The Hamiltonian restricted to each block takes a tridiagonal form. 
The only nonvanishing elements are
\begin{eqnarray} 
   H_{n,n} & = & 4 J + E_{\rm gs} + 2 h n - h L, \nonumber  \\
   H_{n,m} & = & g (1 + e^{ip}), \qquad \, \,\, \,  n-m=1, \nonumber \\
   H_{n,m} & = & g (1 + e^{-ip}), \qquad n-m=-1,
\end{eqnarray}
where $1 \le n,m \le L-1$. Note that the Hermitian conjugate of $H(p)$
is $H(2 \pi -p)$, therefore, for $p \not= 0,\pi$, the spectrum is doubly
degenerate, a result which is not specific of the perturbative
analysis, but holds in general.

To determine the spectrum of $H$, we compute 
\begin{equation}
   D(\ell,{E}) = \det(H - {E} I) \,,
\end{equation}
where $I$ is the identity matrix and $\ell = L-1$ is the dimension
of the matrix associated with $H$.
To simplify the notation, we define 
\begin{equation}
 b = {4 J + E_{gs} + h L\over 2h}, \quad
 z = {2g\over h} \cos(p/2),  \quad  f = 2 h , 
\label{definitions}
\end{equation}
so that $H - {E} I$ takes the form (again we write it explicitly for
$L=4$)
\begin{equation}
\begin{pmatrix} f (b - 3) - {E} & {1\over 2} f z e^{-ip/2} & 0 \\[2mm]
   {1\over 2} f z e^{ip/2}& f (b - 2) - {E} & {1\over 2} f z e^{-ip/2}
  \\[2mm]
     0 & {1\over 2} f z e^{ip/2} & f (b - 1) - {E}
\end{pmatrix} \,.
\end{equation}
For $h = 0$, the determinant and the spectrum of $H$ is easily 
determined (see, e.g., App. B.2 of Ref.~\cite{CPV-15-iswb}).
The energies of  the $n$ levels are: 
\begin{equation}
E_n = E_{gs} + 4 J - 4 g \cos{p\over 2} \cos {\pi n\over L}.
\label{Ekink-h0}
\end{equation}
Here we will use the same methods to obtain an exact result for $h\not=0$. 

By using the properties of the determinant (we expand with respect
to the last row), we obtain the recursion relation:
\begin{eqnarray}
  D(\ell,E) &=& [f(b-1) - {E}]D(\ell-1,{E}+f) \nonumber \\
  & & - {f^2 z^2\over 4} D(\ell-2,{E}+2f) \,,
\label{recursiond}
\end{eqnarray}
which holds for any $\ell \ge 2$, provided we set
$D(1,{E}) = f (b - 1) - {E}$ and $D(0,{E}) = 1$.
To solve it, we fix $\ell$ and define 
\begin{equation}
   a_k = \left({f z \over 2}\right)^{-k} D[k , {E} + (L-1-k) f].
\end{equation}
By replacing $\ell$ with $k+1$ and ${E}$ with ${E}- (k+2-L)f$
in Eq.~(\ref{recursiond}), we obtain 
\begin{equation}
a_{k+1} = {2\over z} (k+c) a_k - a_{k-1}, \quad 
c = b - L + 1 - {E}/f.
\label{recursion-E}
\end{equation}
The solution of the recursion can be obtained by noting that the
Bessel function of the first kind $J_\nu(z)$ and the Neumann-Weber
function (Bessel function of the second kind) $N_\nu(z)$ satisfy a
similar recursion relation. Using Eq.~(8.417) of Ref.~\cite{GradRi}
with $\nu = k+c$ we find
\begin{equation}
   Z_{k+c + 1}(z) = {2\over z} (k+c) Z_{k+c}(z) - Z_{k+c-1}(z)
\label{recBessel}
\end{equation}
with $Z = J$ or $Z = N$.  Thus, we can write the general solution as
\begin{equation}
   a_k = A J_{k+c}(z) + B N_{k+c}(z).
\end{equation}
The coefficients $A$ and $B$ are fixed by the condition that $a_0 = 1$ and  
$a_1 = 2 c/z$. We obtain 
\begin{eqnarray}
  A &=&  -{1\over C} \left[2 c N_{c}(z) - z N_{1+c}(z)\right], \nonumber \\
  B &=&  {1\over C} \left[2 c J_{c}(z) - z J_{1+c}(z)\right], \nonumber \\
  C &=& z \left[J_{c}(z) N_{1+c}(z) - J_{1+c}(z) N_{c}(z)\right].
\end{eqnarray}
Using Eq.~(8.477) of Ref.~\cite{GradRi}, $C = - 2/\pi$.
Moreover, from Eq.~(\ref{recBessel}) with $k = 0$, we finally obtain
\begin{equation}
 A = {\pi z\over 2} N_{-1+c}(z), \quad 
 B = -{\pi z\over 2} J_{-1+c}(z). 
\end{equation}
We have thus an exact expression for $a_k$. Setting $k = \ell = L-1$
we obtain an exact result for the determinant:
\begin{eqnarray}
&& D(\ell,{E}) = {\pi z \over 2} \left({f z \over 2}\right)^{L-1} 
   \times \label{determinant}\\
&& \quad 
   \times \Bigl[N_{-1+c}(z) J_{L-1+c}(z) - J_{-1+c}(z) N_{L-1+c}(z)\Bigr],
\nonumber
\end{eqnarray}
which depends on the energy $E$ only through the variable $c$ defined 
in Eq.~(\ref{recursion-E}).

\subsection{Spectrum} \label{AppA.2}

Using Eq.~(\ref{determinant}) we can obtain the spectrum of $H$. Indeed, 
for given $p, h$, and $L$, we can first compute $z$ and then determine 
the solutions $\nu_n(z)$ of the equation
\begin{equation}
   N_{\nu}(z) J_{\nu+L}(z) - J_{\nu}(z) N_{\nu+L}(z) = 0,
\label{sec-esatta}
\end{equation}
where we set $\nu = c  - 1$. 
The energies are given by
\begin{equation}
{E}_n = E_{gs} + 4 J - h L - 2 h \nu_n(z).
\end{equation}
We now wish to determine the behavior of ${E}_n$ for $L\to\infty$,
$h\to 0$, keeping the ratio $hL$ fixed. We assume $h$ positive,
so that $z \to +\infty$ in the $h \to 0$ limit.  As for $\nu$, we
have numerically verified that the solutions of the secular equation
(\ref{sec-esatta}), also diverge ($\nu_n(z) \to +\infty$) but the
ratio $\nu_n(z)/z$ converges to a constant in the limit.  Under these
conditions, a systematic use of the expansion (8.452) of
Ref.~\cite{GradRi} allows us to prove that the ratio
\begin{equation}
{N_{\nu}(z) \, J_{\nu +L}(z) \over J_{\nu}(z) \, N_{\nu+L}(z)} 
\end{equation}
vanishes exponentially in the infinite-length limit. Thus,
with exponential precision we obtain that the quantities $\nu_n(z)$
satisfy the equation
\begin{equation}
     J_{\nu}(z) = 0.
\end{equation}
Asymptotic results for the Bessel function in the limit of large $\nu$
and $z$ are reported in Refs.~\cite{Olver-54a,Olver-54b,JL-12}.  In
particular, an asymptotic formula for the zeroes is reported in
Ref.~\cite{Olver-54b}, see Eq.~(7.1). To leading order, they can be
obtained by solving
\begin{equation}
   \mathop{\rm Ai}(\nu^{2/3} \zeta) = 0,
\end{equation}
where $\mathop{\rm Ai}(z)$ is the Airy function and $\zeta$ is a
function of $y = z/\nu$. The zeroes of the Airy function all belong to
the negative real axis~\cite{Olver-54b}, and this requires the
function $\zeta$ to be negative which only occurs for $y > 1$.  In
this interval of values of $y$, $\zeta$ is given by
\begin{equation}
\zeta(y) = - \left({3\over2}\right)^{2/3} \left[
   \sqrt{y^2 - 1} - \arccos (1/y)\right]^{2/3}.
\end{equation}
Note that $\zeta(y)$ goes to zero for $y \to 1$ as 
\begin{equation}
    \zeta(y) \approx - 2^{1/3} (y - 1) \left[1 - {3\over 10} (y-1)\right] .
\end{equation}
Thus, if $\alpha_n$ are the zeroes of the Airy function ($\alpha_n < 0$), we 
have 
\begin{equation}
   \nu_n(z) \approx \nu_{{\rm e},n} =  z \,(1 - |\alpha_n| 2^{-1/3} z^{-2/3} ) .
\end{equation}
The approximation becomes exact for $z\to \infty$, since
\begin{equation}
    J_{\nu_{{\rm e},n}} (z) \sim z^{-1}
\end{equation}
in this limit.

\begin{figure}[t]
\includegraphics[width=0.9\columnwidth, clip]{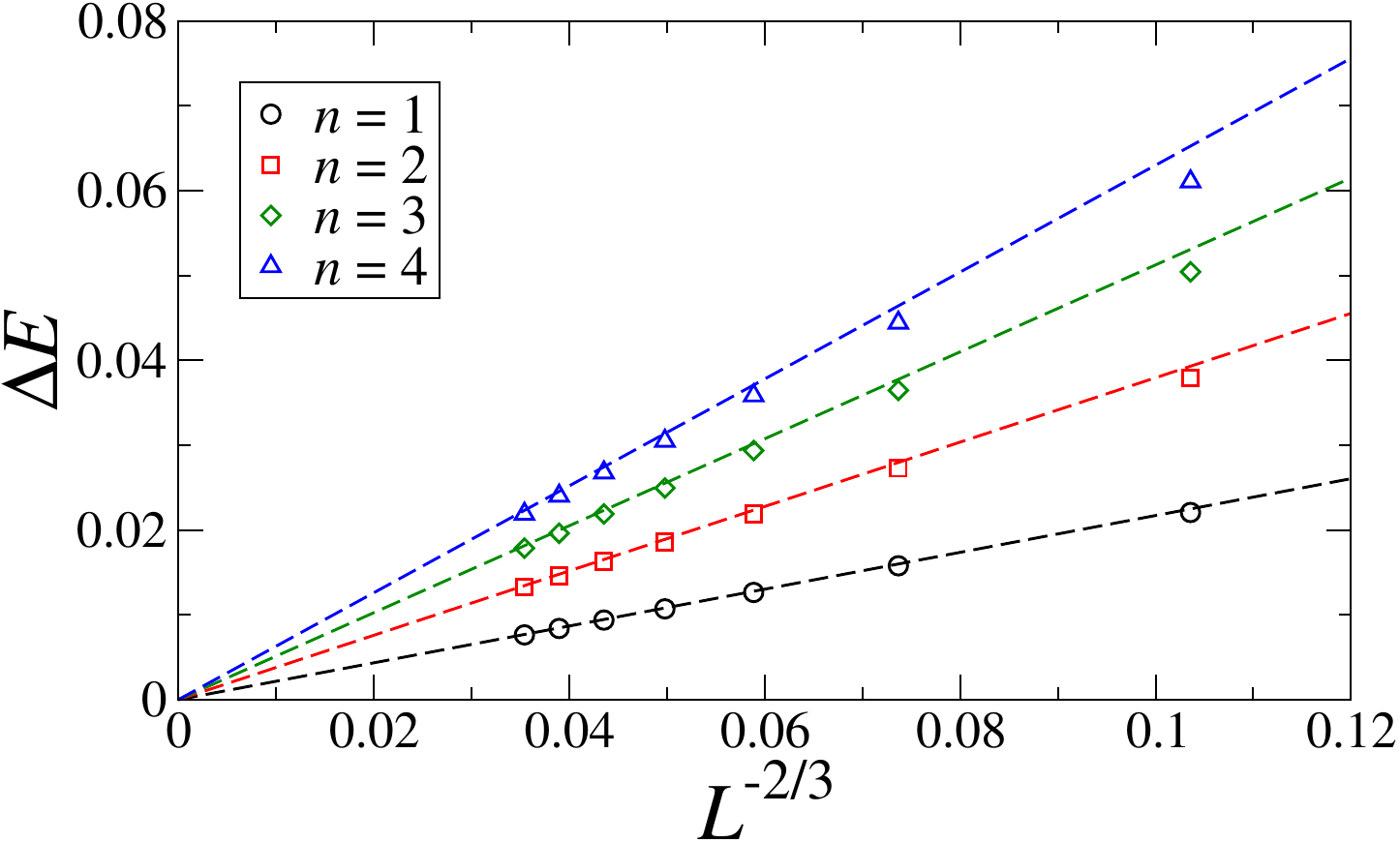}
\caption{Plot of $\Delta E = {E} - E_{gs} - 4 J + h L + 4 g$, for $h L
  = 1/10$, $g = 1/100$, $p=0$, and values of $L$ ranging from 30 to
  150.  We report the four lowest energies.  The straight lines are
  obtained by using the approximate formula (\ref{Enapprox}). }
\end{figure}

Collecting all these results and using the explicit expression for $z$, 
Eq.~(\ref{definitions}),
 we end up with 
\begin{eqnarray}
{E}_n &\approx& E_{gs} + 4 J - h L \\
    &&- 4 g \cos(p/2) \left[
1 - {|\alpha_n|\over 2}  \left
({h \over g \cos(p/2) }\right)^{2/3}\right].
\nonumber 
\end{eqnarray}
Note that the nonanalytic term is of order $L^{-2/3}$ at fixed $h L$.
Corrections are of order $1/L$. Finally, note that $p$ vanishes as
$1/L$. Thus, to order $L^{-2/3}$, there is no $p$ dependence: We
expect the degeneracy in $p$ to be lifted only at order $1/L^2$, as in
the absence of magnetic field. We thus obtain
\begin{equation}
{E}_n = E_{gs} + 4 J - h L - 4 g \left[
1 - {|\alpha_n|\over 2} \left
({h \over g }\right)^{2/3}\right] + O(L^{-1}).
\label{Enapprox}
\end{equation}
It is important to stress that this result only holds in the 
limit $h \to 0$, $L\to \infty$ at fixed $hL$. More precisely, 
it does not hold for $h \to 0$ at fixed $L$, since for finite sizes the 
behavior is analytic in $h$. In this limit, the magnetic-field corrections
are of order $h^2$ at fixed $L$, because of the symmetry under $h\to -h$. 
This analytic behavior should be observed when the magnetic energy, of order
$h L$, is much smaller than the splitting of the levels due to the 
transverse field, which is of the order of $g/L^2$ [more precisely,
$E_n - E_1 \approx 2 \pi^2 g (n^2 - 1)/L^2$, see Eq.~(\ref{Ekink-h0})], i.e.,
for $h \ll g L^{-3}$. Equation~(\ref{Enapprox}) applies instead for 
$h\gg g L^{-3}$. Indeed, if this condition holds, the magnetic 
energy $hL$ is much larger than the correction term of order $g^{1/3} h^{2/3}$.

The expansion depends on the zeroes of the Airy function.
The smallest zeroes correspond to
$\alpha_n = -2.33811, \, -4.08795, \, -5.52056, \, -6.78671, \, -7.94413$
for $n=1,2,3,4,5$. For larger values of $n$, we can use the asymptotic formula
\begin{equation}
\alpha_n \approx -\left[ {3 \pi \over 8} (4 n - 1) \right]^{2/3}.
\label{alphan}
\end{equation}
This is a reasonable approximation even for $n=1$, as it predicts
$\alpha_1 \approx -2.32$, which should be compared with 
the exact result $\alpha_1 = -2.33811\ldots$. 
For the spacing of the levels we thus predict 
\begin{equation}
E_{n+1} - E_n \approx c_n \left({h \over g }\right)^{2/3},
\end{equation}
where $c_n$ decreases with $n$ ($c_n \sim n^{-1/3}$ for $n$ not too small).

\subsection{Magnetization} \label{AppA.3}

\begin{figure}[t]
\includegraphics[width=0.9\columnwidth, clip]{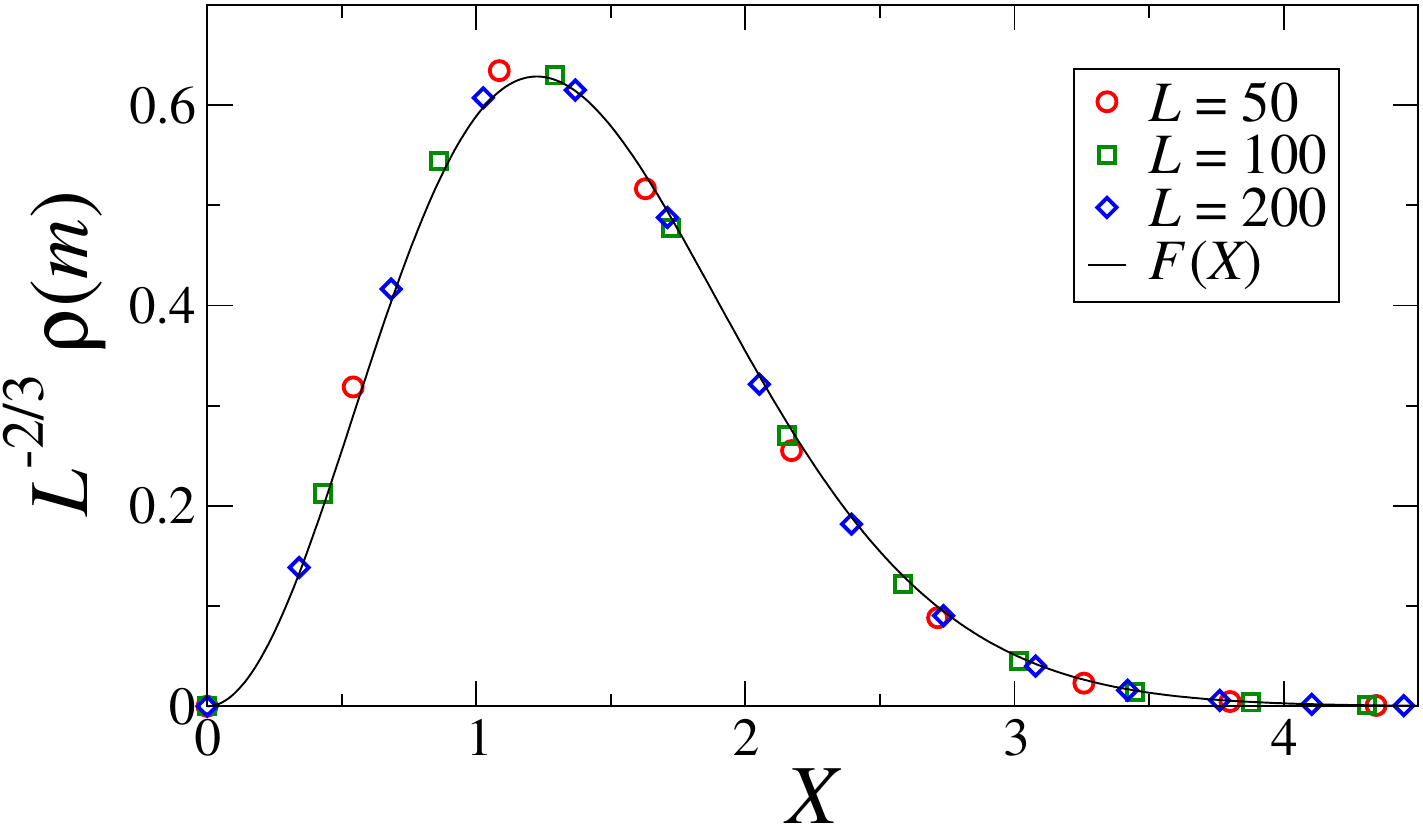}
\caption{Plot of $L^{-2/3} \rho(m)$ versus $X = (1-m)L^{2/3}$, for $h
  L = 1/10$, $g = 1/100$, $p=0$ and three values of $L$.  We consider
  the lowest-energy ($n=1$) kink-antikink state.  We also report the
  scaling furve $F_n(X)$ computed using Eq.~(\ref{fX}).  }
\label{fig:scalingM}
\end{figure}

We now focus on the behavior of the magnetization $M$, defined as
\begin{equation}
  M = {1\over L} \langle \Psi |\sum_x \sigma_x^{(1)} |\Psi\rangle,
\end{equation}
where $|\Psi\rangle$ is a generic normalized state. To determine
its scaling behavior, we define a density function $\rho(m)$.  If
\begin{equation}
  |\Psi\rangle = \sum_i B_i |\Psi_i\rangle,
  \label{bicoeff}
\end{equation}
where $|\Psi_i\rangle$ is an eigenstate of the magnetization operator
with eigenvalue $m_i$, $\rho(m)$ is defined so that
\begin{equation}
  \int_a^b dm\, \rho(m) = \sum_{i:a\le m_i \le b} |B_i|^2,
\end{equation}
for any $-1\le a \le b \le 1$.
The density $\rho(m)$ satisfies 
\begin{equation}
  \int_{-1}^1 dm\, \rho(m) = 1, \qquad
  \int_{-1}^1 dm\, m \rho(m) = M.
\end{equation}
We have determined $\rho(m)$ for the lowest eigenstates of the
kink-antikink Hamiltonian, finding that $\rho(m)$ is strongly peaked
around $m\approx 1$, with a width that goes to zero as
$L\to\infty$. More precisely, it satisfies the scaling law (see
Fig.~\ref{fig:scalingM})
\begin{equation}
  \rho(m) \approx L^{2/3} F(X), \qquad X = (1-m)L^{2/3},
  \label{eq:scalingM}
\end{equation}
which in turn implies 
\begin{equation}
  M \approx 1 - L^{-2/3} \int dX\, X F(X),
\end{equation}
We thus conclude that, in the infinite-size limit, the kink-antikink
states are fully magnetized as the ground state, with corrections of
order $L^{-2/3}$.

To verify Eq.~(\ref{eq:scalingM}) and compute the scaling function
$F(X)$, let us expand the 
eigenstates of the kink-antikink Hamiltonian in terms of the 
vectors $|p,i\rangle$ defined in Eq.~(\ref{App:basis}), 
which are eigenstates of the magnetization operator with eigenvalue
$m_i = 1 - 2 i/L$:
\begin{equation}
|\Psi\rangle = \sum_i B_i \, |p,i\rangle. 
\end{equation}
The coefficients $B_i$ satisfy the recursion relation 
\begin{equation}
e^{ip/2} B_{k-1} + {2\over z} (k + c-1) B_k + 
   e^{-ip/2} B_{k+1} = 0,
\end{equation}
$k = 1,\ldots, L -1$, with boundary conditions $B_0 = B_L = 0$.
This is the analogue of Eq.~(\ref{recursion-E}). The general
solution is
\begin{equation}
B_k = d (-1)^k e^{ipk/2} \left[
   N_{\nu}(z) J_{k+\nu}(z) - J_{\nu}(z) N_{k+\nu}(z)\right],
\end{equation}
which satisfies the condition $B_L = 0$, because of
Eq.~(\ref{sec-esatta}).  Here $d$ is a function of the system
parameters that is fixed by the normalization condition $\langle
\Psi|\Psi\rangle = 1$. As before, we have defined $\nu = c-1$.

The previous result is exact for any value of $L$. For large sizes, we
can use the fact that the secular equation can be simply written as
$J_{\nu}(z) = 0$. Therefore, we obtain
\begin{equation}
B_k = d' (-1)^k e^{ipk/2} J_{k+\nu}(z),
\end{equation}
with $d' = d N_{\nu}(z)$. 
Note that the momentum $p$ appears 
only as a phase, and thus the results for the magnetization depend 
on $p$ only through the variable $z$. 
The function $\rho(m)$ can be estimated by setting 
\begin{equation} 
  \rho(m) = {L\over 2} |B_k |^2 \qquad
\hbox{for} \;\; {(2k - 1)\over L} \le 1-m \le {(2 k + 1) \over L}.
\end{equation}
The scaling function $F(X)$ is then defined as 
\begin{equation} 
   F(X) = \tfrac12 L^{1/3} |B_{XL^{1/3}/2}|^2.
\end{equation}
To evaluate the scaling function, we need the asymptotic 
expansion of $J_{k+\nu}(z)$ for $z\to \infty$ and $\nu = \nu_n(z)$.
Using the results of Refs.~\cite{Olver-54a,Olver-54b,JL-12}, in
the scaling limit we obtain the asymptotic expansion
\begin{equation}
  J_{k+\nu}(z) = 2^{1/3}  z^{-1/3} \mathop{\rm Ai}(\xi_n(X)) [1 + O(z^{-2/3})],
\end{equation}
where 
\begin{eqnarray}
\xi_n(X) &=& \alpha_n + 2^{1/3} z^{-1/3} k \nonumber \\
   &=&  \alpha_n + \left[{hL\over g\cos(p/2)}\right]^{1/3} 
{X\over 2 }.
\end{eqnarray}
As before, since $p\sim 1/L$, only subleading corrections depend on $p$, 
so we can set $p =0$.
The scaling function for the $n^{\rm th}$ kink-antikink state is given by
\begin{equation}
F_n(X) = {1\over 2 N_{1,n}} \left({hL\over g}\right)^{1/3} 
      \mathop{\rm Ai}[\xi_n(X)]^2, 
\label{fX}
\end{equation}
where the prefactor has been determined by requiring $F(X)$ to be normalized.
The constant $N_{1,n}$ is given by 
\begin{equation}
N_{1,n} = \int_{\alpha_n}^\infty dx\, \mathop{\rm Ai}(x)^2; 
\end{equation}
for $n=1$, we have $N_{1,1} \approx  0.491697$. The scaling result (\ref{fX}) 
is reported in Fig.~\ref{fig:scalingM}: the agreement with the numerical 
data is excellent.

To check the previous results, we compare the magnetization computed 
using Eq.~(\ref{fX}) with the expression that follows from the 
Hellmann-Feynman theorem:
\begin{eqnarray}
 {M}_n &=& - {1\over L} {\partial {E}_n\over \partial h}  \nonumber \\
   &=& 1 - {4 |\alpha_n|\over 3} \left({g \over hL}\right)^{1/3} L^{-2/3}.
\label{scalingM2}
\end{eqnarray}
If we use Eq.~(\ref{fX}), we obtain the same scaling behavior. The 
constant $4 |\alpha_n|/3$ is replaced by $N_{2,n}$ where
\begin{equation}
N_{2,n} = {2\over N_{1,n}} 
    \int_{\alpha_n}^\infty dx\, (x - \alpha_n) \mathop{\rm Ai}(x)^2.
\end{equation}
We have verified numerically, 
with 12-digit precision, that indeed
$4 |\alpha_n|/3 = N_{2,n}$, confirming the correctness of the previous 
computation.

Again, we stress that the result (\ref{scalingM2}) only holds in the
limit $h \to 0$, $L\to \infty$ at fixed $hL$. This is
evident from the expression (\ref{scalingM2}) that does not admit a
finite limit for $h\to0$ at fixed $L$.  In the latter case $M_n\sim h$
for small values of $h$. As a final remark, note that $\alpha_n$
scales as $n^{2/3}$ for $n$ not too small, see Eq.~(\ref{alphan}), and
thus the effective length scale that controls the corrections at fixed
$hL$ is $L/n$, implying that larger and larger lattice sizes are
needed to observe the asymptotic behavior of the energy or of the
magnetization of the $n^{\rm th}$ kink-antikink level, as $n$
increases.

\subsection{Comparison with the spectrum of kink-antikink states for 
systems with fixed parallel boundary fields}
\label{AppA.4}

It is interesting to compare the spectrum results for different
boundary conditions. Here we focus on systems with 
parallel boundary fields that force the boundary spins to be parallel.
As before, we consider the restriction of $H$ to the space of
kink-antikink states (the states of energy $E_{\rm gs} + 4 J$ for
$h=g=0$) , which is spanned by the eigenstates $|s_1,\ldots
s_L\rangle$ of $\sigma_x^{(1)}$ with $s_1 = s_L = -1$. Numerically, we
have determined the spectrum of $H_{2k}$ for parallel boundary fields
and periodic boundary conditions. In the case of parallel boundary
fields, we indicate with ${E}_{k,{\rm PBF}}$ the lowest energies of
the spectrum (the energies increase with increasing $k$),
$k=0,\ldots$. In the case of periodic boundary conditions, we indicate
with ${E}_{k,{\rm PBC}}$ the ground-state energies of the system with
$p = 2 \pi k/L$. They indeed represent the lowest energies of the
spectrum as $L$ increases. Note that we are not taking into account
the degeneracy of the levels with $p\not=0$, as this is
related to the symmetry of the system with periodic boundary
conditions under spin inversion, a symmetry which is not present 
for parallel boundary fields. Finally, we consider
\begin{equation}
 \Delta E_k(L) = {E}_{k,{\rm PBF}} - {E}_{k,{\rm PBC}}.
\label{DeltaEk}
\end{equation}
We have studied this quantity for $1\le k\le 10$ for $g = 1/100$ and 
different values of $h L$, considering chains of length $10\le L \le 50$.
In all cases we observe that $\Delta E_k(L)\sim L^{-\beta}$ for $L\to \infty$,
with an exponent that depends on the sign of $hL$.

For $hL=-0.1$, data with $k = 0$ and $1$ are fitted quite precisely
by a power law $\Delta E_k(L) = A L^{-\beta}$, with $\beta \approx
2.01$, and $\beta\approx 2.67$, respectively, which suggest the exact
results $\beta = 2$ and $8/3$ in the two cases.  Results for $hL =
-0.2$ and $-0.02$ are consistent, confirming the estimates of $\beta$.
For $k \ge 2$, data do not appear to be asymptotic and we can only
obtain the lower bound $\beta \gtrsim 1.8$, which would suggest $\beta
= 2$, as for $k = 0$.  For positive values of $hL$---we consider $hL =
0.002,0.1,$ and 0.2---fits of the data with $k=0,1,2$ give estimates
that satisfy $0.6\lesssim \beta\lesssim 0.75$, with significant
corrections that can be taken in into account by assuming $\Delta E_k(L)
= A L^{-\beta} + B L^{-1}$.  These results suggest $\beta = 2/3$ in
all cases.

The previous results allow us to relate the size behavior of $h_1(L)$,
the magnetic field where the lowest-energy kink-antikink state and the
magnetic state $|-\rangle$ have an (avoided) crossing, for periodic
boundary conditions and parallel boundary fields, the case studied in
Ref.~\cite{PRV-18-fowb}. In both cases, Eq.~(\ref{hkbeh}) holds, with
the same constant $a$, as a consequence of the above-reported
analysis: Boundary conditions only affect the next-to-leading
coefficients. Numerical results are in excellent agreement with this
preditions. Indeed, we find $a = 1.0352(1)$ for periodic boundary
conditions, see Sec.~\ref{spectrum}, to be compared with $a=1.0370(5)$
reported in Ref.~\cite{PRV-18-fowb}.

\end{document}